\RequirePackage{fix-cm}
\documentclass[smallcondensed]{svjour3}     
\smartqed  
\usepackage{graphicx}
 \usepackage{natbib}
 \usepackage[hidelinks]{hyperref}
 \usepackage[caption=false]{subfig}
 \usepackage{makecell}
 \usepackage[ruled,linesnumbered]{algorithm2e}
 \usepackage{mathtools}
 \usepackage[misc]{ifsym}
 \usepackage{multirow}
 \usepackage[dvipsnames]{xcolor}
 \usepackage{appendix}

\begin{document}

\title{Embed2Detect: Temporally Clustered Embedded Words for Event Detection in Social Media
}

\titlerunning{Embed2Detect}        

\author{Hansi Hettiarachchi \textsuperscript{$\ddag$} \and \\
        Mariam Adedoyin-Olowe \textsuperscript{$\ddag$} \and \\
        Jagdev Bhogal \textsuperscript{$\ddag$} \and \\
        Mohamed Medhat Gaber \textsuperscript{$\ddag$}
}

\authorrunning{H. Hettiarachchi et al.} 

\institute{\Letter \hspace*{1mm} H. Hettiarachchi \at \hspace*{4mm} \email{Hansi.Hettiarachchi@mail.bcu.ac.uk} 
\and \hspace*{4mm}
M. Adedoyin-Olowe \at \hspace*{4mm} \email{Mariam.Adedoyin-Olowe@bcu.ac.uk} 
\and \hspace*{4mm}
J. Bhogal \at \hspace*{4mm} \email{Jagdev.Bhogal@bcu.ac.uk}
\and \hspace*{4mm}
M. M. Gaber \at \hspace*{4mm} \email{Mohamed.Gaber@bcu.ac.uk}
\and 
\textsuperscript{$\ddag$} \hspace*{2mm} School of Computing and Digital Technology, Birmingham City University, UK}

\date{Received: date / Accepted: date}

\maketitle

\begin{abstract}
Social media is becoming a primary medium to discuss what is happening around the world. Therefore, the data generated by social media platforms contain rich information which describes the ongoing events. Further, the timeliness associated with these data is capable of facilitating immediate insights. However, considering the dynamic nature and high volume of data production in social media data streams, it is impractical to filter the events manually and therefore, automated event detection mechanisms are invaluable to the community. Apart from a few notable exceptions, most previous research on automated event detection have focused only on statistical and syntactical features in data and lacked the involvement of underlying semantics which are important for effective information retrieval from text since they represent the connections between words and their meanings. In this paper, we propose a novel method termed \textbf{Embed2Detect} for event detection in social media by combining the characteristics in word embeddings and hierarchical agglomerative clustering. The adoption of word embeddings gives \textbf{Embed2Detect} the capability to incorporate powerful semantical features into event detection and overcome a major limitation inherent in previous approaches. We experimented our method on two recent real social media data sets which represent the sports and political domain and also compared the results to several state-of-the-art methods. The obtained results show that \textbf{Embed2Detect} is capable of effective and efficient event detection and it outperforms the recent event detection methods. For the sports data set, Embed2Detect achieved 27\% higher F-measure than the best-performed baseline and for the political data set, it was an increase of 29\%. 

\end{abstract}

\vspace{-2mm}
\section*{Declarations}
\textbf{Funding:} Not applicable\\
\textbf{Conflicts of interest/Competing interests:} No conflicts\\
\textbf{Availability of data and material:} The data sets used for the experiments of this research are published on \\  \url{https://github.com/hhansi/twitter-event-data-2019} \\
\textbf{Code availability:} The implementation of suggested system is available on \url{https://github.com/hhansi/embed2detect}

\section{Introduction}
\label{sec:intro}
Social media services such as Twitter and Facebook are becoming increasingly popular. A recent survey by \citet{dave2019global} estimated the number of active social media users around the world in January 2019 as 3.484 billion; 45\% of the total population. The average of the global increase in social media usage since January 2018 was found to be 9\%. Another analysis was conducted on active users on social media in July 2019 to rank social media services based on popularity \citep{clement2019global}. According to its results, the majority of the services have millions of users with Facebook leading with a user base of 2,375 million. Approximately 511,200 tweets per minute were recorded in 2018 \citep{james2019data}. 

The data produced on social media contain different information such as opinions, breaking news and personal updates. Also, social media facilitates fast information dispersal because of its large user base which covers a vast geographical area \citep{castillo2011information}. In some cases, social media was found to broadcast news faster than traditional news media by an analysis which compared Twitter trending topics with CNN news headlines \citep{kwak2010twitter}. Due to the inclusion of diverse information and real-time propagation to large groups, nowadays, there is a high tendency to consider social media as information networks which provide newsworthy contents. In 2017, the proportion of American adults who got news from social media was found to be 67\% \citep{Gottfried2017NewsUA}. Consequently, news services such as BBC and CNN also use social media actively to instantly publish news to a huge user base. Nonetheless, it is impractical to analyse the data manually to extract important or newsworthy content from social media, because of its huge volume and dynamic nature. Therefore, in order to utilise the social media data effectively, the requirement of an automated and accurate event detection method is crucial \citep{small2014review}.

Considering this requirement, different methods were suggested by previous research for event detection in social media as further discussed in Section \ref{sec:related-work}. Apart from a few notable exceptions, most of the methods were based on statistical and syntactical features in data and lacked the involvement of semantical features. A language is mainly built using two phenomena, namely, syntax and semantics \citep{sag1987information}. Syntax defines the arrangement of words in word sequences, and semantics describes the connections between words and their meanings. Using both syntax and semantics, languages allow different word sequences to express the same idea. This impact is widely demonstrated in the social media text, due to the diversity in users.  For example, consider the tweets:

\begin{quote}
    \textit{`There are 13 million people living in poverty in the UK. 13M!!!  Yet some MPs will vote for the deal with NO impact assessments. That 13M could become 20M?!\#VoteTheDealDown \#PeoplesVoteMarch \ \#PeoplesVote \#StopBrexit'}
\end{quote}
\begin{quote}
    \textit{`Luciana Berger - Steve Barclay confirmed that no economic analysis of the \#BrexitDeal has been done... let that sink in. So how can we be expected to vote on a deal, that will affect this country for decades, today? \ \#VoteDownTheDeal \#PeoplesVote'}
\end{quote}

\noindent which were posted during the Brexit Super Saturday 2019. Even though both tweets describe the same idea, there are no common words between them except for a few hashtags. In addition, different word phrases such as \textit{`impact assessments'} and \textit{`economic analysis'} were used to mention the same subject discussed in them. In such cases, semantics are needed to understand the relationships between terms to extract valuable information. 

Focusing the importance of semantics to natural language processing (NLP), word embeddings such as Word2Vec \citep{mikolov2013efficient} with high capability in preserving syntactic and semantic relationships between words were introduced. These embeddings were successfully used within many NLP related tasks such as news recommendation \citep{zhang2019dynamic}, question classification \citep{yilmaz2020deep} and community detection \citep{vskrlj2020embedding} recently. Similarly, Word2Vec embeddings were used for an event detection approach named $W_{E}C$ too \citep{comito2019word}. $W_{E}C$ is mainly based on pre-trained word embeddings and online document clustering. However, pre-trained models will be less effective in the context of social media due to the availability of modified or misspelt words. The inability to recognise such word can lead to a serious information loss. 

Considering the lack of semantic involvement in previous methods, this research proposes a novel event detection method termed \emph{Embed2Detect} which combines the characteristics of word embeddings and hierarchical agglomerative clustering. Rather than using pre-trained word embeddings, we propose to use self-learned word embeddings which can capture the characteristics specific to the targeted corpus. Also, without relying on direct clusters, we consider the temporal variations between clusters and vocabularies using a time-based sliding window model to identify event occurrences. We targeted a token-based clustering approach to successfully handle the scenarios where a single document contains multiple event details. In addition to considering the underlying syntax and semantics, event detection also requires the incorporation of statistical features in the text, to measure the qualities of events such as popularity. In our method, this requirement also fulfilled by using self-learned word embeddings and vocabulary change measures, which capture statistics. In summary, Embed2Detect is an improved method, which considers all the important features in textual data; syntax, semantics and statistics, which are needed for effective event detection.

To evaluate the proposed method, two recent social media data sets which represent two diverse domains; sports (English Premier League 19/20 on 20 October 2019 between the teams: Manchester United Football Club (FC) and Liverpool FC) and politics (Brexit Super Saturday 2019) are used. We used Twitter to collect the data because it is widely considered as an information network than social media \citep{adedoyin2016rule, kwak2010twitter}, and has limited restrictions with enough data coverage for this research. To measure the performance, we used the evaluation metrics of recall, precision, F-measure and keyword recall, which are widely used to evaluate the event detection methods. Further, we compared the effectiveness and efficiency of our method with three recently proposed event detection methods as baselines. We obtained promising results for the evaluation which outperformed the baseline results. 

To the best of our knowledge, Embed2Detect is the only method which uses self-learned word embedding-based temporal cluster similarity for event detection in social media. In summary, we list the contributions of this paper as follows:
\color{black}
\begin{itemize}
    \item 
    Proposing a novel method named Embed2Detect for event detection in social media by involving the semantical features using the self-learned word embeddings in addition to the statistical and syntactical features in the text;
    \item leveraging self-learned word embeddings for more effective and flexible event detection which is independent of characteristics specific to the social media service, language or domain; 
    \item the application of Embed2Detect to recent real data sets to provide an insight on effectiveness and universality of the method with a comparison over state-of-the-art methods;
    \item the publication of recent social media data sets\footnote{Data sets are available on   \url{https://github.com/hhansi/twitter-event-data-2019}} 
    which represent different domains (i.e. sports and politics) with ground truth event labels to support other research in the area of event detection; and
    \item the release of method implementation as an open-source project\footnote{Embed2Detect implementation is available on
    \url{https://github.com/hhansi/embed2detect}} 
    to support applications and research in the area of event detection. 
\end{itemize}

The rest of this paper is organised as follows. Available methods for event detection in social media and their capabilities are discussed in Section \ref{sec:related-work}. Section \ref{sec:background} describes the background details including the support of word embeddings and hierarchical clustering for this research. The problem addressed by this research is stated in Section \ref{sec:problem-definition} and the proposed approach is explained under Section \ref{sec:suggested-approach}. Following this, a comprehensive experimental study is available under Section \ref{sec:experiment-results}. Finally, the paper is concluded with a discussion in Section \ref{sec:conclusion}.

\section{Related work}
\label{sec:related-work}
Considering the importance of automatic event detection in social media, different methods have been proposed by previous research with the association of different techniques and characteristics including graph theory, rule mining, clustering, burstiness and social aspect. These techniques were supported by different text representations including tokens, n-grams, vectors, etc.; and extracted keywords such as named entities, noun phrases and hashtags as further discussed below. Additionally, more comprehensive surveys done by \cite{weiler2017survey} and \cite{hasan2018survey} can be referred to obtain a deep understanding of qualities and discrimination of available methods.

\vspace{-3mm}
\paragraph{\textbf{Graph Theory:}} Following the successful application of graph theory in sociology and social network analysis, there has been a tendency to use graph-based solutions for event detection in social media.   \citet{sayyadi2009event} proposed to transfer a data stream into a KeyGraph, which represents the keywords by nodes and connects the nodes if corresponding keywords co-occurred in a document so that the communities in the graph represent the events that occurred in the data stream. As keywords, noun phrases and named entities with high document frequency were considered, and for community detection, betweenness centrality score was used. Later research suggested using Structural Clustering Algorithm for Networks (SCAN) to extract the communities in the graph and social media posts as graph nodes \citep{schinas2015multimodal}. Unlike the betweenness centrality-based cluster detection, SCAN has the ability to recognise bridges of clusters (hubs) and outliers, to allow for the sharing of hubs between clusters and recognition of outliers as noise \citep{xu2007scan}. However, this approach is unable to handle the events which are unknown to the training data, as it uses a supervised learning technique to identify similar posts during the graph generation. Other recent research suggested a named entity-based method considering the high computational cost associated with graph generation \citep{edouard2017graph}. After identifying the entities, only the context around them was considered to extract nodes, edges and weights. Even though keyword-based methods speed up the graph processing, they are less expandable due to the usage of language or domain-specific features for keyword extraction.

\vspace{-3mm}
\paragraph{\textbf{Rule Mining:}} Previous research showed a trend of applying rule mining techniques for event detection. Based on Association Rule Mining (ARM), \citet{adedoyin2013trcm} proposed a method for temporal analysis of evolving concepts in Twitter which was named Transaction-based Rule Change Mining (TRCM). To generate the association rules, hashtags in tweets were considered as keywords. This event detection methodology was further evolved by showing that specific tweet change patterns, namely, unexpected and emerging, have a high impact on describing underlying events \citep{adedoyin2016rule}. Having a fixed support value for Frequent Pattern Mining (FPM) was found to be inappropriate for dynamic data streams and it was solved by the dynamic support calculation method proposed by \citet{alkhamees2016event}. FPM considers all terms in equal utility. But, due to the short length in social media documents, frequency of a specific term related to an event could increase rapidly compared to other terms. Based on this finding, \citet{choi2019emerging} suggested High Utility Pattern Mining (HUPM) which finds not only the frequent but also the high in utility itemsets. In this research, the utility of terms was defined based on the growth rate in frequency. Even though the latter two approaches which are based on FPM and HUPM are not limited to the processing of special keywords such as hashtags, they are focused on only identifying the topics/events discussed during a period without recognising temporal event occurrence details. 

\vspace{-3mm}
\paragraph{\textbf{Clustering:}} By considering the dynamicity and unpredictability of social media data streams, there has been a tendency to use clustering for event detection. \citet{mccreadie2013scalable} and \citet{nur2015combination} showed that K-means clustering can be successfully used to extract events. In order to improve efficiency and effectiveness, they clustered low dimensional document vectors, which were generated using Locality Sensitive Hashing (LSH) and Singular Value Decomposition (SVD), respectively. Considering the requirement of predefining the number of events in K-means clustering, there was a motivation for hierarchical or incremental clustering approaches \citep{corney2014spot,li2017real,nguyen2019hot,morabia2019sedtwik}. Different data representations such as word n-grams \citep{corney2014spot}, semantic classes \citep{li2017real} and segments \citep{morabia2019sedtwik} were used with hierarchical clustering. Rather than focusing on individual textual components, \cite{comito2019bursty} suggested to cluster social objects which combine the different features available within documents. This idea was further improved with the incorporation of word embeddings \citep{comito2019word}. In addition to word embeddings, term frequency-inverse document frequency (tf-idf) vectors also used to represent documents during clustering \citep{nguyen2019hot, hasan2019real}. The document clustering-based approaches assume that a single document only belongs to the same event. But with the increased character limits by most of the social media services, multiple events can be described within a single document. Considering this possibility, individual components (e.g. n-grams, segments) are more appropriate to use with clustering. Among the above-mentioned such data representations, segments are more informative, specific and easy to extract, because they are separated using semantic resources such as Wikipedia.

\vspace{-3mm}
\paragraph{\textbf{Burstiness:}} In communication streams, a burst is defined as a transmission which involves a larger amount of data than usual over a short time. \citet{van2012automatic} suggested that occurrences of sports events in Twitter can be recognised by analysing the bursts of tweets in the data stream. Similarly, \cite{li2014online} proposed an incremental temporal topic model which recognises the bursty events based on unusual tweet count changes. But the events which do not lead to any notable increase in the data volume would be missed if only the data at peak volumes are considered. To overcome this limitation, another research proposed to use bursts in word n-grams \citep{corney2014spot}. This research argues that even when the data volume is stable, there will be an increase in word phrases specific to a particular event. However, frequency-based measures cannot differentiate the events from general topics such as car, music, food, etc., because social media contains a large proportion of data relevant to these topics. Moreover, the bursts in frequency will appear when an event becomes more popular or is trending. To overcome these issues, bursts in word acceleration was suggested by other research \citep{xie2016topicsketch}. Using the acceleration, events could be identified more accurately at their early stages. 

\vspace{-3mm}
\paragraph{\textbf{Social Aspect:}} Recently, there was a focus on the social aspect considering the impact the user community has on events. \citet{guille2015event} proposed an approach which focuses on the bursts in mentions to incorporate the social aspect of Twitter for event detection. Since the mentions are links added intentionally to connect a user with a discussion or dynamically during re-tweeting, the social aspect of data can be revealed using them. Proving the importance of the social aspect, this method outperformed the methods which are only based on term frequency and content similarity \citep{benhardus2013streaming, parikh2013events}. Recent research has also suggested an improved version of Twevent \citep{li2012twevent} by integrating more user diversity-based measures: retweet count and follower count with segment burstiness calculation \citep{morabia2019sedtwik}. However, the measures which gauge the social aspect are mostly specific to the social media platform and the incorporation of them would require customising the main flow accordingly. 

Considering the textual features used in the above-mentioned event detection approaches, it is clear to us that the majority of previous research mainly focused on statistical features (e.g. term frequency, tf-idf, or burstiness), and syntactical features (e.g. co-occurrence, or local sensitivity). As a subdomain of information retrieval from text, effective event detection in social media also requires the proper inclusion of semantical features even though we could only find a few methods which considered the underlying semantics as described in Section \ref{sub-sec:usage-semantics}.

\subsection{Usage of semantics in event detection} \label{sub-sec:usage-semantics}
When we closely analysed how semantics is used by previous research for event detection in social media, we found some rule-based and contextual prediction-based approaches as further discussed below. 

\citet{li2017real} defined an event as a composition of answers to WH questions (i.e. who, what, when and where). Based on this definition, they considered only the terms which belong to the semantic classes: proper noun, hashtag, location, mention, common noun and verb for their event detection method. Rule-based approaches were used for the term extraction and categorisation. Likewise, another recent research \citep{nguyen2019hot} also used a rule-based approach to extract the named entities in the text in order to support their event detection method. Using the named entities, documents and clusters were represented as entity-document and entity-cluster inverted indices which were used for candidate cluster generation. Both of these methods only categorised the terms into semantical groups for the recognition of important terms related to events. Thus, none of these methods has the ability to identify the connections between words.

In contrast to the rule-based approaches, \citet{chen2017online} suggested a deep neural network-based approach for event detection. To identify event-related tweets a neural network model was used and to input the data into the network tweets were converted into fixed-length vectors using pre-trained GloVe embeddings \citep{pennington2014glove}, while capturing the semantic and syntactic regularities in the text. It is not appropriate to use supervised learning techniques for real-time event detection, because they require prior knowledge of events which can vary due to the dynamic nature in data streams and event-specific qualities. Similarly, \cite{comito2019word} proposed to use tweet representations generated using pre-trained Skip-gram embeddings \citep{mikolov2013efficient}. This method was based on incremental clustering which is more suitable for event detection in social media than supervised learning. However, both of these methods use pre-trained embeddings which unable to capture the characteristics specific to the targeted corpus such as modified or misspelt words.

In summary, based on the available literature, we could not find any event detection approach which adequately involves semantics of the underlying text considering the characteristics specific to social media. We propose our approach with the intention to fill this gap for more effective event identification. 

\section{Background}
\label{sec:background}
Considering the limitations in available approaches for event detection, we adopt an approach which is based on word embeddings and hierarchical clustering in this research. The background details for word embeddings and their capabilities are discussed in Section \ref{sub-sec:word-embeddings}. The basic concepts of hierarchical clustering are explained in Section \ref{sub-sec:hierarchical-clustering}.

\subsection{Word embeddings} \label{sub-sec:word-embeddings}

Word embeddings are numerical representations of text in vector space. Depending on the learning method, they are categorised into two main groups as frequency-based and prediction-based embeddings. Frequency-based embeddings consider different measures of word frequencies to represent text as vectors while preserving statistical features. Unlike them, prediction-based embeddings learn representations based on contextual predictions while preserving both syntactical and semantical relationships between words. Considering these characteristics, we focus on prediction-based word embeddings in this research and will use the term `word embeddings' to refer to them. 

Different model architectures such as Neural Network Language Model (NNLM) \citep{bengio2003neural} and Recurrent Neural Network Language Model (RNNLM) \citep{mikolov2010recurrent} were proposed by previous research for the generation of word embeddings based on contextual predictions. However, considering the complexity associated with them, log-linear models which are known as Word2vec models \citep{mikolov2013efficient} were suggested and popularly used with NLP applications. There are two architectures proposed under Word2vec models: (1) Continuous Bag-of-Words (CBOW) and (2) Continuous Skip-gram. CBOW predicts a word based on its context. In contrast to this, Skip-gram predicts the context of a given word. According to the results obtained by model evaluations, \citet{mikolov2013efficient} showed that these vectors have a high capability in preserving syntactic and semantic relationships between words.

Among the Word2Vec algorithms, we focus on Skip-gram model in this research, because it resulted in high semantic accuracy compared to CBOW \citep{mikolov2013efficient, mikolov2013distributed}. Also, based on the initial experiments and analyses, Skip-gram outperformed the CBOW model. More details of the Skip-gram architecture are described in Section \ref{sub-sec:skip-gram-model}. Following this theoretic exposure, Section \ref{sub-sec:skip-gram-vector-spaces} discusses the qualities of word embeddings obtained by training Skip-gram models on real data sets which are useful for event detection. 

\subsubsection{Skip-gram model} \label{sub-sec:skip-gram-model}
Skip-gram model is a log-linear classifier which is composed by a 3-layer neural network with the objective to predict context/surrounding words of a centre word given a sequence of training words $w_{1}, w_{2}, ... w_{n}$ \citep{mikolov2013distributed}. More formally, it focuses on maximizing the average log probability of context words $w_{k+j}|-m \leq j \leq m, j \neq 0$  of the centre word $w_{k}$ by following the objective function in Equation \ref{eq:sg-obtective-function}. The length of the training context is represented by $m$.

\begin{equation} \label{eq:sg-obtective-function}
j = \frac{1}{n} \sum_{k=1}^{n} \sum_{-m \leq j \leq m, j \neq 0} \log p (w_{k+j}|w_ {k}) 
\end{equation}

The probability of a context word given the centre word; $p (w_{k+j}|w_ {k})$ is computed using the softmax function. 

\begin{equation} \label{eq:softmax-probability}
    p (w_{o}|w_{i})= \frac{\exp ({v_{w_{o}}^{'}}^{T} v_{w_{i}})}{\sum_{w=1}^{N} \exp ({v_{w}^{'}}^{T} v_{w_{i}})}
\end{equation}

\begin{figure*}
\centering
  \includegraphics[width=0.55\textwidth]{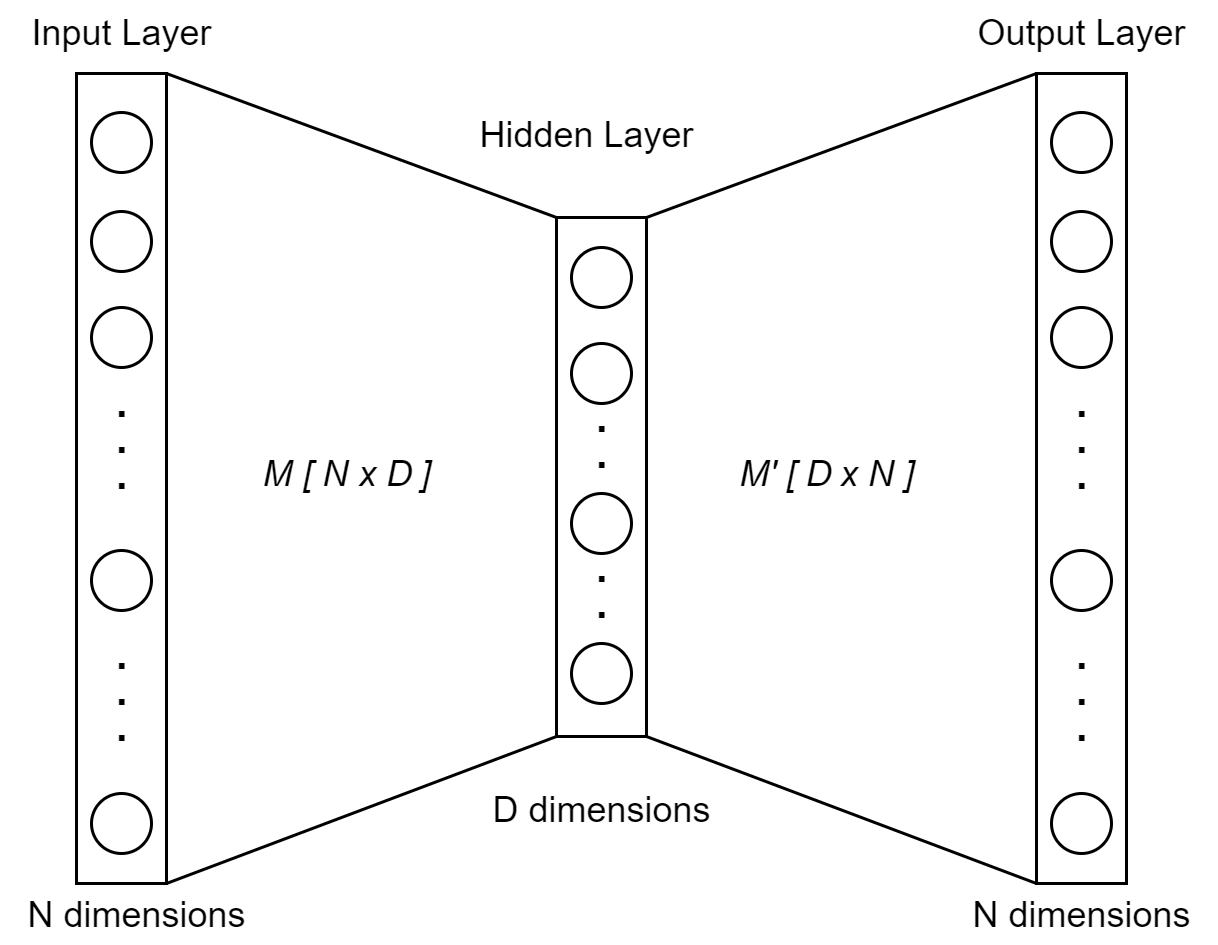}
  \caption{Architecture of Skip-gram model}
\label{fig:skipgram-architecture}  
\end{figure*}

In Equation \ref{eq:softmax-probability}, $w_{o}$ and $w_{i}$ represent the output and input (i.e. context and centre words) and $N$ represents the length of vocabulary. The input and output vectors of a word $w$ is represented by $v_{w}$ and $v_{w}^{'}$. The input vectors for words are taken from input-hidden layer weight matrix $M$ which is sized $N \times D$ where $D$ is the number of hidden layers. Likewise, output vectors are taken from hidden-output layer weight matrix $M^{'}$ which is sized $D \times N$. The architecture of Skip-gram model including weight matrices is shown in Figure \ref{fig:skipgram-architecture}.

Once the model converges, it obtains an ability to predict the probability distributions of context words with good accuracy. At that point, instead of using the model for trained purpose, adjusted weights between the input and hidden layers will be extracted as word representations or embeddings. Thus, by changing the number of hidden layers of the model, the number of neurons and also the dimensionality of vectors can be changed. Following the training procedure, model weights are adjusted by learning the connections between nearby words. Provided a sufficient data corpus, learning the connections between nearby words allows capturing underlying syntax and semantics with the capability of grouping similar words more effectively. 

\subsubsection{Skip-gram vector spaces learned on event data} \label{sub-sec:skip-gram-vector-spaces}
An event discussed in a data stream will result in a collection of documents which describe that event using a set of words related to it. Due to the learning based on contextual predictions, word embeddings has an ability to locate the vectors of contextually closer words in nearby vector space or group similar words. This characteristic allows generating nearby vectors for the event-related words when the embeddings are learned on the corresponding document corpus. 

Let us consider the sample events mentioned in Table \ref{tab:sample-events} which are  extracted from English Premier League 19/20 on 20 October 2019 between the teams Manchester United FC and Liverpool FC relating to the players Marcus Rashford and Roberto Firmino. Both events corresponding to Firmino are about missed attempts. Rashford has two different events relating to a foul and a goal. By analysing the Twitter data posted during each minute, we could find a significant amount of tweets which discuss these events. In these tweets, foul related words were used in the context of word \textit{`Rashford'} at 16:52 and goal-related words were used at 17:06. Likewise, missed attempt related words were used in the context of \textit{`Firmino'} at 16:40 and 17:04.

\begin{table}
\caption{Sample events occurred during English Premier League 19/20 on 20 October 2019 (Manchester United - Liverpool)}
\label{tab:sample-events}       
\begin{tabular}{lll}
\hline\noalign{\smallskip}
Time & Event & Description  \\
\noalign{\smallskip}\hline\noalign{\smallskip}
16:40 & Attempt missed & Attempt by Roberto Firmino (Liverpool) \\
16:52 & Foul & \makecell[l]{Foul by Marcus Rashford (Manchester United) on Virgil van\\ Dijk (Liverpool)} \\
17:04 & Attempt saved &  Attempt by Roberto Firmino (Liverpool) \\
17:06 & Goal & First goal by Marcus Rashford (Manchester United)\\
\noalign{\smallskip}\hline
\end{tabular}
\end{table}

\begin{figure*}
\centering
  \includegraphics[width=0.95\textwidth]{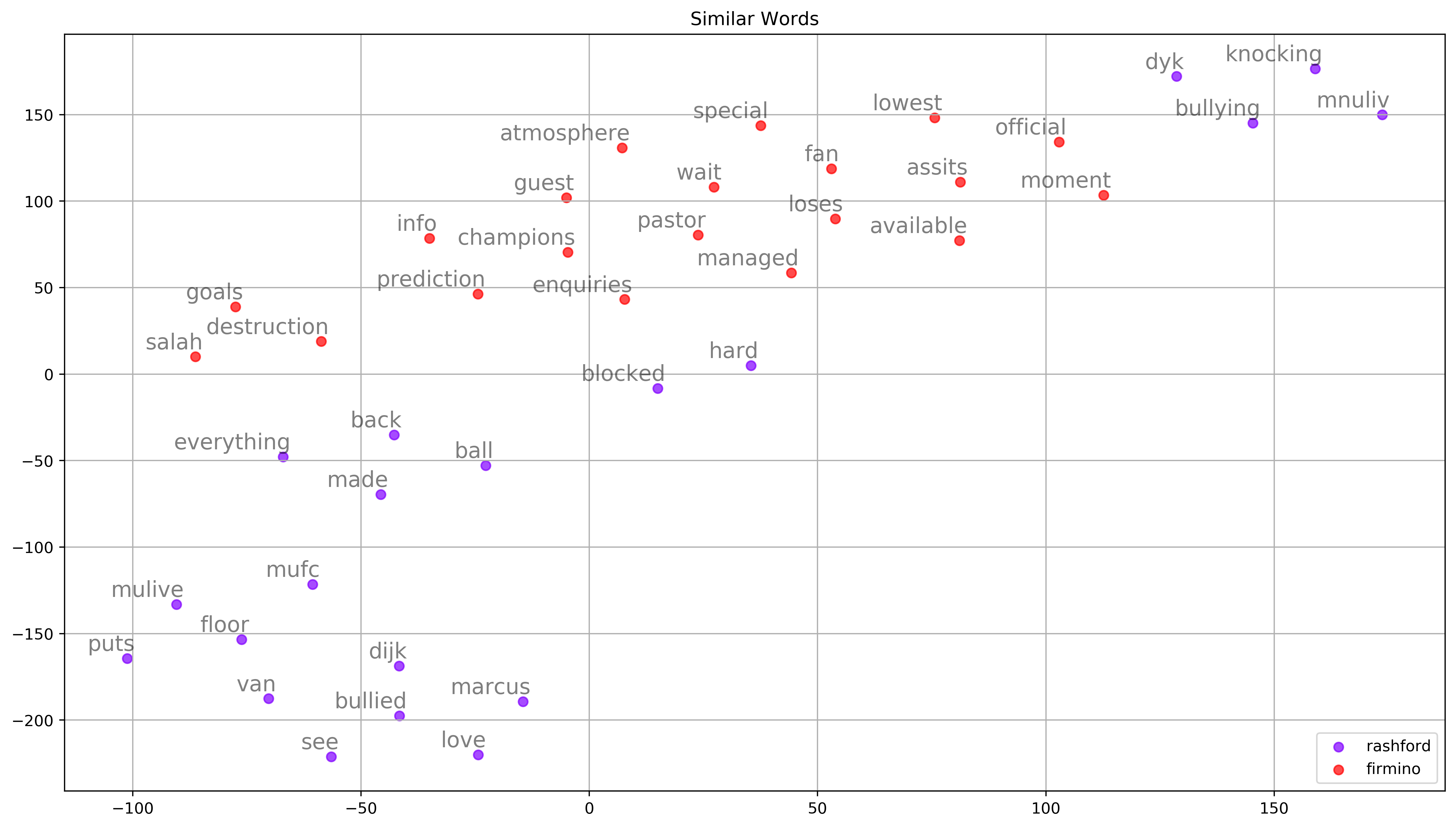}
  \caption{t-SNE visualisation of tokens closer to the words; \textit{`Rashford'} and \textit{`Firmino'}} within time window 2019-10-20 16:52 - 16:54
\label{fig:similar-words1}       
\end{figure*}

\begin{figure*}
\centering
  \includegraphics[width=0.95\textwidth]{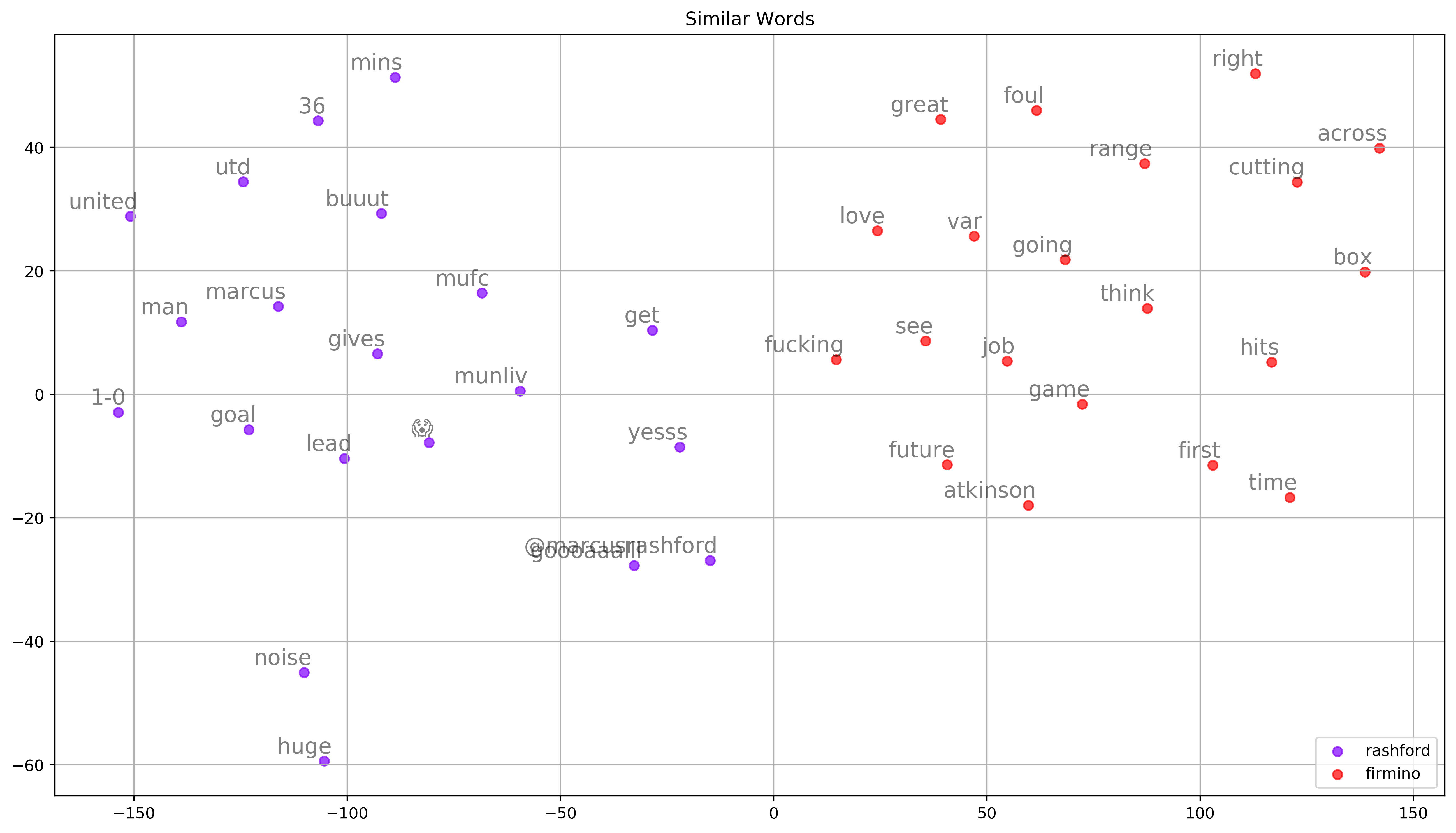}
  \caption{t-SNE visualisation of tokens closer to the words; \textit{`Rashford'} and \textit{`Firmino'}} within time window 2019-10-20 17:06 - 17:08
\label{fig:similar-words2}       
\end{figure*}

To analyse the word embedding distribution over vector space and its temporal variations relating to these events, we trained separate Skip-gram models for each time window using Twitter data. In order to provide enough data for embedding learning, 2-minute time windows were used. Using the learned embeddings, most similar words to the player names Rashford and Firmino were analysed during the time windows 16:52-16:54 and 17:06-17:08. To visualise the similar words in a two-dimensional plane, T-distributed Stochastic Neighbor Embedding (t-SNE) algorithm \citep{maaten2008visualizing} was used and resulted graphs are shown in Figures \ref{fig:similar-words1} and \ref{fig:similar-words2}. 

The similar word visualisation during 16:52-16:54 (Figure \ref{fig:similar-words1}) shows that the foul related words are located closer to the word \textit{`Rashford'} in the vector space. Also, after 12 minutes, few words related to the missed attempt at 16:40 such as \textit{`loses'} and \textit{`destruction'} can be seen closer to the word \textit{`Firmino'}. But, during 17:06-17:08, we can see more words related to the saved attempt as nearby vectors to \textit{`Firmino'}, because this event occurred 2 minutes back (Figure \ref{fig:similar-words2}). Also, the goal scored during 17:06 can be clearly identified by the words closer to \textit{`Rashford'}. This time window has clearly separated nearby vector groups for  \textit{`Firmino'} and \textit{`Rashford'} compared to the previous window 16:52-16:54 to indicate that both events are actively discussed during this time because they happened recently. 

These similar word analyses prove that nearby vector groups have the ability to represent the events. Thus, the events described in a document corpus can be identified using the learned embeddings.  Skip-gram word embeddings locate directly as well as indirectly related words to an event in closer vector groups. For an example, the top 20 similar words to \textit{`Rashford'} at the time window; 17:06-17:08 (Figure \ref{fig:similar-words2}), contains the words such as \textit{`goal', `1-0', `mufc'} and \textit{`36'} which are directly related to the event \textit{goal scored at 36 minute}. Also, similar words contain words such as \textit{`huge'} and \textit{`noise'} which relate indirectly to the event but describe it more. These characteristics associated with Skip-gram word embeddings can be utilised for effective event detection in social media data.

\subsection{Hierarchical clustering} \label{sub-sec:hierarchical-clustering}
Even though flat clustering (e.g. K-means) is efficient compared to hierarchical clustering, flat clustering requires the number of clusters to be predefined. Considering the unpredictability associated with social media data, it is not practical to identify the number of events in advance. Therefore, hierarchical clustering is more appropriate for social media data streams. Another advantage in hierarchical clustering is the output of a hierarchy or structure of data points, which is known as dendrogram rather than just returning the flat clusters. This hierarchy can be used to identify connections between data points. Considering these advantages,   hierarchical clustering is used for our event detection approach.

There are two types of hierarchical clustering algorithms, bottom-up or agglomerative and top-down or divisive \citep{manning2008introduction}. In hierarchical agglomerative clustering (HAC), all data points are considered as separate clusters at the beginning and then merge them based on cluster distance using a linkage method. The commonly used linkage criteria are single, complete and average. In single linkage, the maximum similarity is considered and in complete linkage, the minimum similarity is considered. Average of all similarities are considered in the average linkage. In contrast to HAC, hierarchical divisive clustering (HDC), considers all data points as one cluster at the beginning and then divide them until each data point is in its own cluster. For data division, HDC requires a flat clustering algorithm. 

HDC is more complex compared to HAC, due to the requirement of a second flat clustering algorithm. Therefore, when processing big data sets,  HDC is recommended to use with some stopping rules to avoid the generation of complete dendrogram in order to reduce the complexity \citep{roux2018comparative}. Since we need to process big data sets and focus on clusters as well as complete dendrograms, we decided to use HAC for this research.

\section{Problem definition}
\label{sec:problem-definition}
The problem targeted by this research is automatically detecting events in (near) real-time from social media data streams. The concept behind a data stream is introduced with Definition \ref{def-data-stream}.

\begin{definition} \label{def-data-stream}
    \textit{Social Media Data Stream: } A continuous and chronological series of posts or documents $d_{1}, d_{2},... d_{i}, d_{i+1},...$ generated by social media users. 
\end{definition}

Looking at available event detection approaches, they can be mainly divided into two categories from the perspective of the input data stream, as general and focused. In the general scenario, the whole data stream is processed \citep{mccreadie2013scalable, nguyen2019hot}. In the focused scenario, a user-centred data stream filtered from the whole data stream is processed. Two types of filtering techniques as keyword-based and location-based were commonly used. In keyword-based filtering, a domain-specific data stream will be extracted using a set of keywords \citep{aiello2013sensing, alkhamees2016event, comito2019word}. In location-based filtering, a data stream composed by a set of documents posted by users in a particular location will be extracted \citep{li2012twevent, guille2015event}. Comparing the two filtering techniques, the location-based method seems to add unnatural restrictions, because events of a corresponding location can be reported by users who are located elsewhere and location details are not available with all user accounts. 

Rather than focusing on the whole data stream which is a combination of multiple domains, we address a keyword-based user-centred scenario in this research. This approach was selected considering the restrictions in other filtering techniques and real-world requirements on information extraction. In many real scenarios, people or domain experts need the quick extraction of information in an interested domain rather than extracting all the information available \citep{aiello2013sensing}. For examples, football fans would like to know football updates, fire brigades would like to know fire updates and BBC politics news crew would like to know political updates. 

In this setup, the whole stream needs to be narrow downed initially using some keywords (seed terms) specific to the targeted domain. In the case of social media data streams, they can be easily filtered by commonly used tags in the area of interest. In the majority of cases, these tags will be known by domain experts. Also, many applications can identify trending tags in social media corresponding to different domains\footnote{Popular hashtags under different domains can be found at \url{http://best-hashtags.com/}}. Definition \ref{def-filtered-stream} introduces the concept of the keyword-based filtered data stream.  

\begin{definition} \label{def-filtered-stream}
    \textit{Filtered Data Stream: } A filtered or narrow downed data stream which consists of posts that contain at least one of the selected seed terms. 
\end{definition}

Events were described using various definitions by previous research. \citet{sayyadi2009event} defined an event as some news related thing happening at a specific place and time. Also, events were considered as occurrences which have the ability to create an observable change in a particular context \citep{aldhaheri2017event}. Focusing on the content of events, another research described an event as a composition of answers to WH questions (i.e. who, what, when and where) \citep{li2017real}. Considering the main idea used to describe an event, we use the Definition \ref{def-event} for events. 

\begin{definition} \label{def-event}
    \textit{Event: } An incident or activity which happened at a certain time and discussed or reported significantly in social media. 
\end{definition}

Additionally, we use the concept of time windows to formulate the targeted problem as it is widely used in previous research \citep{aiello2013sensing, adedoyin2016rule, morabia2019sedtwik}. Given a filtered data stream, it will be separated into time windows so that we can assess each window to identify event occurred time windows (Definition \ref{def-event-window}). The length of time windows needs to be provided by domain experts considering the intended update rate. For highly evolving domains like football, time window length needs to be short enough to capture the quick updates and for slowly evolving domains like politics, time window length need to be large enough to capture slowly developed updates. Similarly, the control of event significance also needs to be given to the domain experts, because we cannot define a fixed significance level for different people groups and domains. 

\begin{definition} \label{def-event-window}
    \textit{Event Occurred Time Window: } Duration of time, where at least one event has occurred. 
\end{definition}

In summary, the aim of the system described in this paper is, given a filtered data stream, identifying event occurred time windows in (near) real-time including the corresponding event-related words. Using the hyper-parameters, users are allowed to set the significance and update rate of interested events.

\subsection{Notations of terms}
Providing that the proposed approach is time window-based, the notations $W_{t}$ and $W_{t+1}$ are used to denote two consecutive time windows at time $t$ and $t+1$. All the notations which are commonly used throughout this paper are summarised in Table \ref{tab:notations}.

\begin{table}
\caption{Summary of notations used in the paper}
\label{tab:notations}       
\begin{tabular}{lll}
\hline\noalign{\smallskip}
Notation & Description  \\
\noalign{\smallskip}\hline\noalign{\smallskip}
$W_{t}$ & window at time t \\
$W_{t+1}$ & window at time t+1 (consecutive time window to $W_{t}$)\\
$d_{i}$ & document i in a data stream \\
$w_{i}$ & word/token i in a data corpus \\
$v_{i}$ & word embedding corresponding to the word/token i; $w_{i}$ \\
$vocab_{t}$ & vocabulary corresponding to the data at $W_{t}$ \\
$vocab_{t+1}$ & vocabulary corresponding to the data at $W_{t+1}$ \\
$N$ & length of the vocabulary \\
$dl$ & dendrogram level \\
$dl_{(w_{i},w_{j})}$ & number of shared dendrogram levels between tokens; $w_{i}$ and $w_{j}$ from root \\
$dl_{r\rightarrow x}$ & number of dendrogram levels from root; $r$ to node; $x$ \\
$L$ & set of leaf nodes in a dendrogram \\
\noalign{\smallskip}\hline
\end{tabular}
\end{table}

\section{Embed2Detect}
\label{sec:suggested-approach}
\begin{figure*}
\centering
  \includegraphics[width=0.65\textwidth]{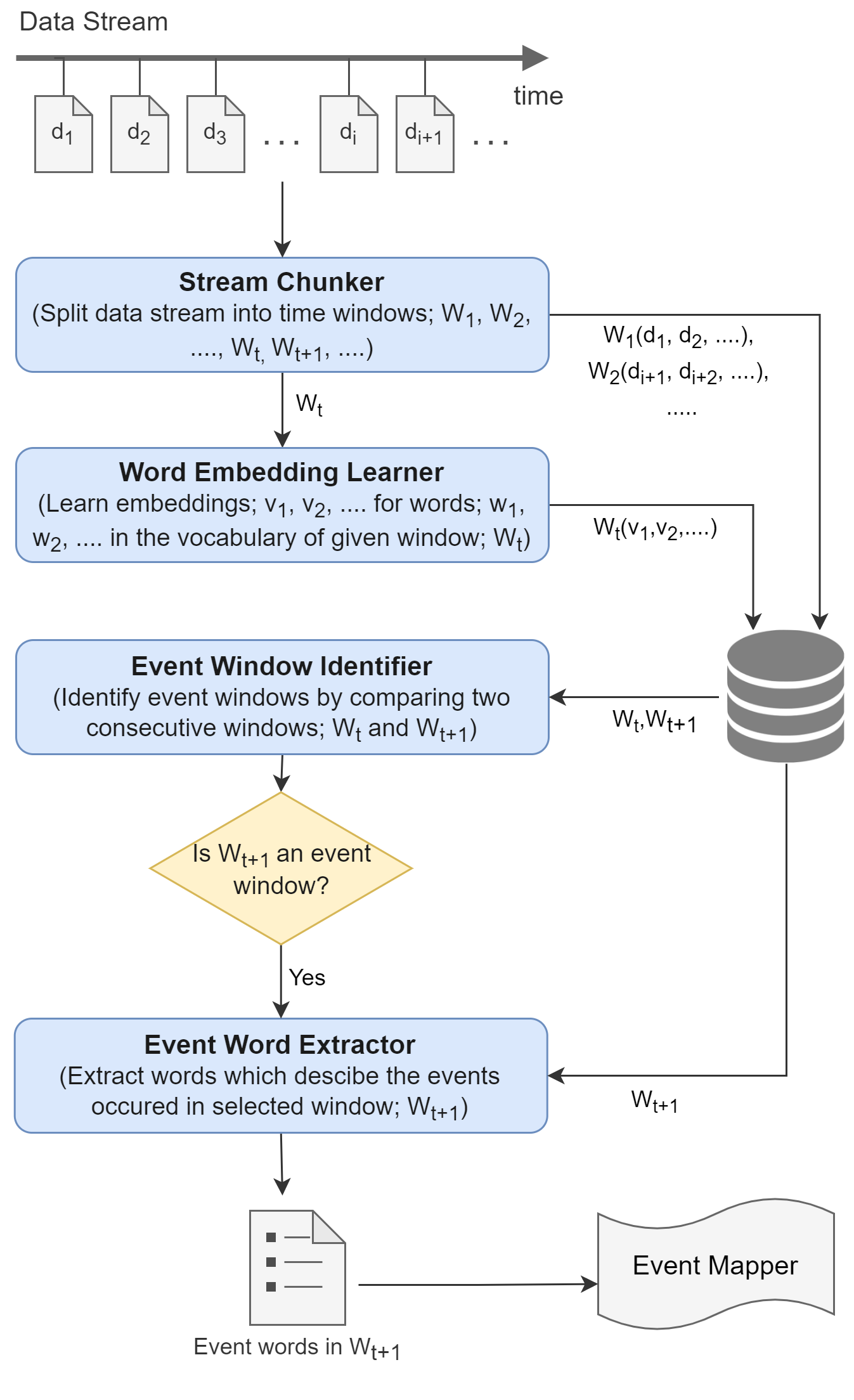}
  \caption{Overview of proposed method for event detection; Embed2Detect}
\label{fig:method-overview}       
\end{figure*}

As Embed2Detect, we propose an event detection approach which is based on word embeddings and hierarchical agglomerative clustering. The main novelty of this approach is the involvement of corpus oriented semantical features for event detection using self-learned word embeddings. Further, the temporal variations between clusters and vocabularies are considered to identify events without relying on clusters directly. The Embed2Detect system contains four main components: (1) stream chunker, (2) word embedding learner, (3) event window identifier and (4) event word extractor as shown in Figure \ref{fig:method-overview}. Self-learned word embeddings are used during event window identification and event word extraction phases. In order to evaluate the performance of this approach, event mapper is used to map detected events with ground truth events during experiments. Each of the components is further described in Sections \ref{sub-sec:stream-chunker} - \ref{sub-sec:event-word-extractor}. Finally, the computational complexity of Embed2Detect is discussed in Section \ref{sub-sec:computational-complexity}.


\subsection{Stream chunker} \label{sub-sec:stream-chunker}
Data stream mining is supported by three different time models, namely, landmark model, tilted-window model and sliding window model \citep{tsai2009mining}. In the landmark model, all the data from a specific time to present is considered equally. The tilted-window model treats recent data with high importance compared to old data. Sliding window model splits the data stream into windows based on a fixed time period or number of transactions and performs data mining tasks on the data that belong to each window. 

Among these models, the time-based sliding window model was widely used by previous research work in event detection \citep{sayyadi2009event, alkhamees2016event, adedoyin2016rule, choi2019emerging}. Analysing the performance of previous methods and considering the requirement of temporal event identification, Embed2Detect also uses the sliding window model with a fixed time frame for event detection in social media data streams. 

Stream chunker is the component which facilitates the separation of the data stream into windows. Depending on the evolution of events which need to be identified, the length of time frames can be adjusted. Smaller time frames are preferred for highly evolving events.

\subsection{Word embedding learner} \label{sub-sec:word-embedding-learner}
In order to incorporate statistical, syntactical and semantical features in text for event detection, word embeddings are used. Without using pre-trained word embeddings, this research proposes to learn embeddings on the targeted corpus to capture its unique characteristics such as modified or misspelt words and emoticons. The word embedding learner transfers the text in social media posts in a selected time window to a vector space. For each time window, different vector spaces are learned to capture variations between them. Learned word embedding models are stored to facilitate event window identification and event word extraction. 

Considering the high-quality vector representations by the Skip-gram algorithm, we used it to learn embeddings in Embed2Detect. Due to the simplicity in this model architecture and usage of small training corpora (chunks of a data stream), time complexity on learning is not considerably high to make a bad impact on real-time event detection. 

\subsection{Event window identifier} \label{sub-sec:event-window-identifier}
Given a chronological stream of time windows $W_{1},W_{2},...W_{t},W_{t+1},...$ , event window identifier recognises the windows where events have occurred. Since an event is an incident or activity which happened and discussed, such occurrence should make a significant change in data in the corresponding time window compared to its previous window. Based on this assumption, our method identifies windows with higher change than a predefined threshold ($\alpha$) as event windows. From the perspective of use cases, this threshold mainly defines the significance of targeted incidents. A low $\alpha$ value would identify less important events too. Since normalised values are used to measure the change, possible values of $\alpha$ are ranged between 0 and 1.

Before moving into the change calculation phase, we preprocess the text in social media documents for more effective results and efficient calculations. We do not conduct any preprocessing steps before learning the word embeddings except tokenizing to preserve all valuable information, which would help the neural network model to figure things out during word embedding learning. As preprocessing in event window identification phase, redundant punctuation marks and stop words in the text are removed.  Further tokens with a frequency below a predefined threshold ( $\beta$ ) are removed as outlier tokens (e.g. words which are misspelt, or used to describe non-event information).

\begin{figure*}
\centering
  \includegraphics[width=0.65\textwidth]{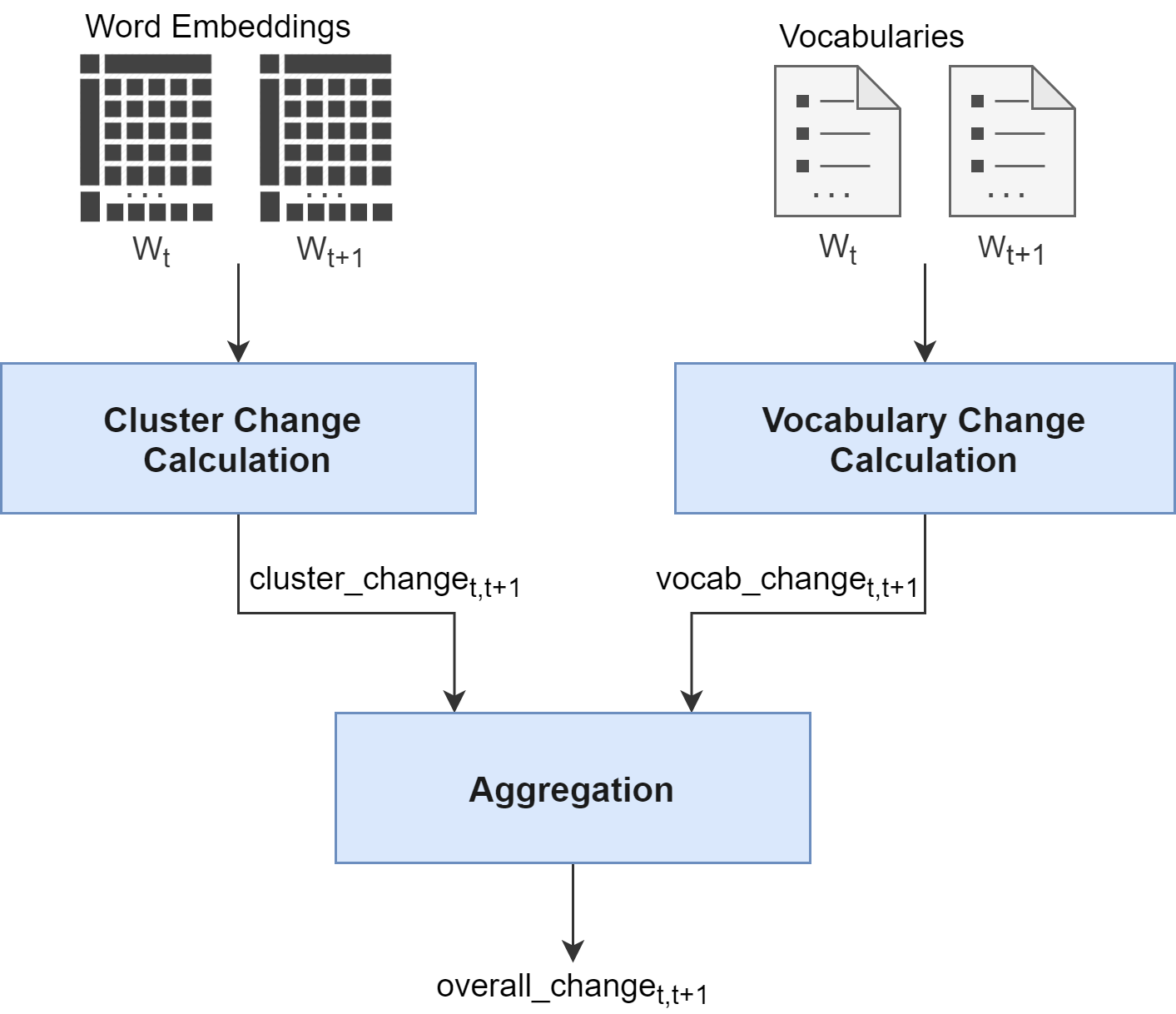}
  \caption{Overview of window change calculation}
\label{fig:change-calculation-overview}       
\end{figure*}

An event occurrence will make changes in nearby words of a selected word or introduce new words to the vocabulary over time. For example, in a football match, if a goal is scored at $W_{t}$, \textit{`goal'} will be highly mentioned in the textual context of a player's name. If that player receives a yellow card unexpectedly in $W_{t+1}$, new words; \textit{`yellow card'} will be added to the vocabulary and they will appear in the context of the player's name, except the word \textit{`goal'}. Following these findings, in Embed2Detect, we consider two measures which capture the changes in nearby words and vocabularies to calculate textual data change between two consecutive time windows $W_{t}$ and $W_{t+1}$. To compute the nearby word changes, we propose a measure which is based on clustered word embeddings under cluster change calculation (Section \ref{sub-sec:cluster-change-calculation}). According to the Section \ref{sub-sec:skip-gram-vector-spaces}, word embeddings can be used effectively to identify nearby word changes based on both syntactical and semantical aspects. Vocabulary change calculation is described in Section \ref{sub-sec:vocabulary-change-calculation}. The final value for the overall textual change between time windows is calculated by aggregating the two measures, cluster change and vocabulary change. As the aggregation method, we experimented maximum and average value calculations (Section \ref{sub-sec:aggregation-method}). Comparing these two methods, the best results could be obtained by using the maximum calculation. An overview for window change calculation is shown in Figure \ref{fig:change-calculation-overview} and complete flow of event window identification is summarised in Algorithm \ref{algo:event-window-identification}.

\begin{algorithm}
\label{algo:event-window-identification}
\caption{Event Window Identification}
\SetAlgoLined
\SetKwInOut{Input}{input}
\KwResult{$eventWindows$: time windows where events occurred}
 $eventWindows$ = []\;
 $\alpha$ = Predefined threshold for overall data change\;
 $Windows$ = Array of time windows \;
 \For{index 1 to length(W)-1}{
  $W_{t}$ =  $Windows[index]$\;
  $W_{t+1}$ = $Windows[index+1]$\;
  $vocab_{t}$ = vocabulary at index\;
  $vocab_{t+1}$ = vocabulary at index+1\;
  \text{/* Measure cluster change */}\;
  $commonVocab$ = common vocabulary for $vocab_{t}$ and $vocab_{t+1}$ \;
  $N$ = Length of $commonVocab$ \;
  $matrix_{t}$ = Similarity matrix at t using $commonVocab$ \;
  $matrix_{t+1}$ = Similarity matrix at t+1 using $commonVocab$ \;
  $diffMatrix$ = $|matrix_{t+1}-matrix_{t}|$ \;
  \text{/** Get average on upper triangular matrix **/} \;
  $clusterChange$ =$\sum_{i=1}^{N} \sum_{j=i+1}^{N} diffMatrix[i,j] / ((N \times(N-1))/2)$ \;
  \text{/* Measure vocabulary change */}\;
  $vocabChange$ = $|vocab_{t+1}-vocab_{t}|/|vocab_{t+1}|$\;
  \text{/* Measure overall change */}\;
  $overallChange$ = $max(clusterChange, vocabChange)$\;
  \If{$overallChange \geq \alpha$}{
  $eventWindows.Add(W_{t+1})$\;
  }
 }
\end{algorithm}

\subsubsection{Cluster change calculation} \label{sub-sec:cluster-change-calculation}
Cluster change calculation is proposed to measure nearby word or word group changes over time. To facilitate this calculation, separate clusterings need to be generated per each time window $W_t$. We propose to cluster tokens that include words as well as other useful symbols such as emojis. As token representations, self-learned word embeddings are used while preserving syntactical and semantical features of the underlying corpus. According to previous research, document clustering approaches were more popularly used with event detection \citep{nur2015combination, nguyen2019hot, comito2019bursty, comito2019word}. However, with the recent increments made to character limits by social media services (e.g. Twitter increased 140 character limit to 280), there is a possibility to contain multiple event details in a single document. Therefore, the token level is more appropriate than the document level to identify events. In addition to limiting only to words, useful symbols such as emojis were incorporated as they are widely used to express ideas in social media. 

As the clustering algorithm, we chose hierarchical agglomerative clustering (HAC) considering its advantages and the tendency by previous research (Section \ref{sub-sec:hierarchical-clustering}). As the linkage method, we used the average scheme in order to involve all the elements that belong to clusters during distance calculation. In average linkage, the distance between two clusters; $C_{i}$ and $C_{j}$ is measured by following the Equation \ref{eq:average-linkage} \citep{mullner2011modern},

\begin{equation} \label{eq:average-linkage}
    D(C_{i},C_{j}) = \frac{1}{|C_{i}||C_{j}|} \sum_{w_{p} \in C_{i}} \sum_{w_{q} \in C_{j}} d(w_{p},w_{q}),
\end{equation}

\noindent where $d(w_{p},w_{q})$ represents the distance between cluster elements $w_{p}$ and $w_{q}$ which belong to the clusters $C_{i}$ and $C_{j}$ respectively. This distance is measured using cosine distance, because it proved effectiveness for measurements in textual data \citep{mikolov2013efficient, mikolov2013distributed, antoniak2018evaluating}. Since cosine distance calculation is independent from magnitude of vectors, it does not get biased by the frequency of words \citep{schakel2015measuring}.

Even though we use hierarchical clustering, our method does not rely on direct clusters, as the majority of available methods \citep{li2017real, comito2019word}. We propose to use temporal cluster changes measured over time windows using dendrograms to identify nearby word changes. In this setting, the requirement for a cluster threshold can be eliminated. This elimination is advantageous in the context of social media because it is hard to define a static threshold suitable for all event clusters considering their diversity. 

After generating clusters, a matrix-based approach is used to compute cluster change. Token similarity matrices are generated for each time window $W_{t}$ considering its next time window $W_{t+1}$. The token similarity matrix is a square matrix of size $N \times N$ where $N$ is the number of tokens in the vocabulary. Each cell in the matrix $matrix[i,j]$ represents the cluster similarity between $token_{i}$ and $token_{j}$. To calculate the cluster similarity between tokens, we propose dendrogram level (DL) similarity measure (Section \ref{sub-sec:dendrogram-level-similarity}) which is based on hierarchical clusters of token embeddings. To compare similarity matrices between two consecutive time windows, a common vocabulary is used for matrix generation. Since we compare $W_{t+1}$ against $W_{t}$, preprocessed vocabulary at $t+1$ $vocab_{t+1}$ is used as the common vocabulary for both windows. After generating the similarity matrices at $t$ and $t+1$ using DL similarity between tokens, the absolute difference of matrices is calculated. Then the average on absolute differences is measured as the value for cluster change in $W_{t+1}$ compared to $W_{t}$.  During the average calculation, we only considered the values at the upper triangular matrix except the diagonal, because the matrix is symmetric around the diagonal.

\subsubsection{Dendrogram level (DL) similarity} \label{sub-sec:dendrogram-level-similarity}
\begin{figure*}
\centering
  \includegraphics[width=0.6\textwidth]{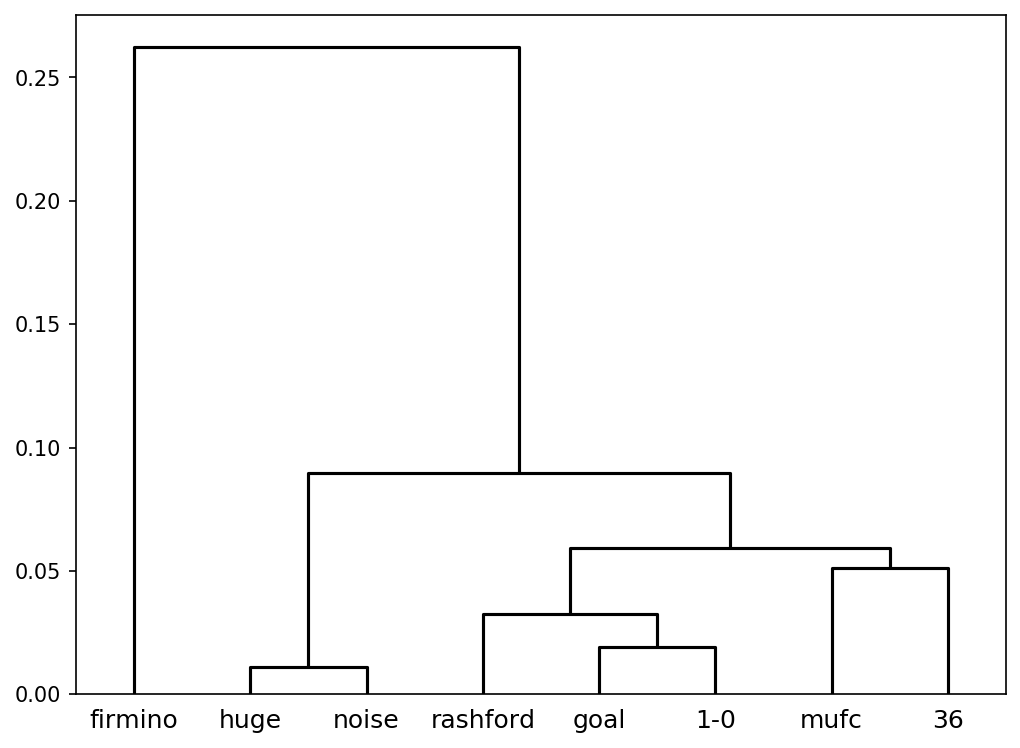}
  \caption{Sample dendrogram (y-coordinate denotes the cosine distance and x-coordinate denotes the selected words)}
\label{fig:sample-dendrogram}       
\end{figure*}

A dendrogram is a tree diagram which illustrates the relationships between objects. These diagrams are typically used to visualise hierarchical clusterings. A sample dendrogram generated on a selected word set from tweets posted during the first goal of English Premier League 19/20 on 20 October 2019, between Manchester United and Liverpool is shown in Figure \ref{fig:sample-dendrogram}. Each merge happens considering the distance between clusters of words and they are represented by horizontal lines. Merges between the closer groups such as the name of the player who scored the goal \textit{`rashford'} and cluster which contains the word \textit{`goal'} happen at low distances ($\approx 0.025$). In contrast to this, merges between distant groups such as another player name \textit{`firmino'} and cluster of \textit{`goal'} happen at high distance values ($\approx 0.25$). Similarly, a dendrogram built on a corpus from any domain preserves informative relationships expressed in the corpus.  

Focusing on the characteristics associated with dendrograms, we suggest the dendrogram level (DL) similarity to measure token similarity based on their cluster variations. Each horizontal line or merge represents a dendrogram level. Given a dendrogram, the similarity between a word pair $w_{i}$ and $w_{j}$ is calculated as the normalised value of shared levels from root between those two words, as follows.

\begin{equation} \label{eq:dl-similarity}
DL\ Similarity_{(w_{i},w_{j})} = \frac{dl_{(w_{i},w_{j})}}{max(dl_{r\rightarrow x}: x \in L) + 1}
\end{equation}

The numerator of Equation \ref{eq:dl-similarity} represents the number of shared dendrogram levels between $w_{i}$ and $w_{j}$ from the root. The denominator represents the maximum number of levels between root and leaf nodes. We added leaf node level also as a separate level during maximum level count calculation to make sure only the similarity between the same token is 1 ($DL\ Similarity_{(w_{i},w_{i})} = 1$). For example, the maximum number of dendrogram levels from root to leaves in the diagram in Figure \ref{fig:sample-dendrogram} is 5. By adding the leaf node level, the maximum level count becomes 6. The count of shared levels between words \textit{`rashford'} and \textit{`goal'} is 4. But, words; \textit{`firmino'} and \textit{`goal'} shares only 1 level, because they appear in distant clusters. In measures, DL similarities between these words are as follows.

\begin{gather*}
    DL\ Similarity_{(rashford,goal)} = \frac{4}{6} = 0.667\\
    DL\ Similarity_{(firmino,goal)} = \frac{1}{6} = 0.167
\end{gather*}

\subsubsection{Vocabulary change calculation} \label{sub-sec:vocabulary-change-calculation}
A vocabulary is a set of distinct words that belong to a particular language, person, corpus, etc. In this research, we consider the tokens that belong to data corpora at each time window as separate vocabularies. Vocabulary change calculation is proposed to measure new word addition into time windows over time. Also, it incorporates the statistical details in the data set. In order to have a comparable value over all time windows, we calculated normalised vocabulary change value for $W_{t+1}$ compared to $W_{t}$ following the Equation \ref{eq:vocab-change}.

\begin{equation} \label{eq:vocab-change}
Vocabulary\ Change_{(t,t+1)} = \frac{|vocab_{t+1} - vocab_{t}|}{|vocab_{t+1}|}
\end{equation}

The numerator of Equation \ref{eq:vocab-change} represents the cardinality of new tokens that appeared in the vocabulary of $W_{t+1}$ compared to $W_{t}$, and the denominator represents the size of the vocabulary that belongs to $W_{t+1}$.

\subsection{Event word extractor} \label{sub-sec:event-word-extractor}
After identifying a time window as an event occurring window, event word extractor facilitates the extraction of words in that window which are related to the occurred events. Since events make changes to the textual corpus, this component marks all the words in a window $W_{t+1}$ which showed cluster changes compared to its previous windows $W_{t}$ as event words. Since we use a common vocabulary between consecutive windows during similarity matrix generation, cluster change calculation identifies the newly added words to $W_{t+1}$ also as words with changes. All words belong to the word pairs with $DL similarity$ change above 0 are considered as the words which have temporal cluster changes. 

\subsection{Computational complexity} \label{sub-sec:computational-complexity}
Analysing the components of Embed2Detect architecture, word embedding learner and event window identifier are the most computationally complex components available. Compared to them, the complexity of stream chunker and event word extractor is negligible. Therefore, for time and space complexity calculations, we only consider word embedding learner and event window identifier. 

The training complexity of Skip-gram architecture is proportional to $C\times(D+D\times\log{N})$, where $C$ is the maximum distance of the words, $D$ is the dimensionality of vectors and $N$ is the size of vocabulary \citep{mikolov2013efficient}. Under event window identifier, there are two complex sub-components, clustering and similarity matrix generation. For $N$ sized vocabulary, the time complexity for HAC algorithm is $O(N^{2}\log{N})$ \citep{manning2008introduction} and for matrix generation is $O(N^{2})$. By combining all these chained components, the time complexity of Embed2Detect can be calculated as $O(CD\log{N} + N^{2}\log{N})$. For the used application, both $C$ and $D$ values are comparatively smaller than $N$. Therefore, the time complexity can be further simplified to $O(N^{2}\log{N})$.

Following the 3-layer architecture, the space requirement of  Skip-gram model is equivalent to $N\times D + D\times N$. Similarly, both HAC algorithm and similarity matrix generation have a space complexity of $O(N^2)$. Considering all cost values, the total space complexity of Embed2Detect can be summarised as $O(DN + N^2)$. Using the same assumption mentioned above, this can be simplified to $O(N^2)$.

Based on the complexities summarised above, the vocabulary size $N$ has a high impact on the computational complexity of this approach. According to the recent reports, approximately 511,200 tweets per minute were recorded in 2019 \citep{james2019data}. For a 30-minute time window, the total tweet count would be approximately 15M. As the time window length, 30 minutes is selected as a sufficiently long period to highlight the worst-case scenario. Looking at available twitter-based word embedding models, the Word2Vec\_Twitter model is trained on 400M tweets and has a vocabulary of 3M tokens \citep{godin2015multimedia}. Another popularly used model, GloVe Twitter is trained on 2B tweets and has a vocabulary of 1.2M tokens\footnote{GloVe pre-trained model details are available on \url{https://nlp.stanford.edu/projects/glove/}}. Focusing on the worst-case scenario, if there were 3M vocabulary for 400M tweets, $N$ can be approximated to 0.1M (100,000) for 15M tweets. Since our approach is targeted in processing a filtered data stream specific to a particular domain, $N$ should be notably smaller than this value approximation (0.1M) in a real scenario. Further, the size of vocabulary can be controlled using the frequency threshold ($\beta$) mentioned in Section \ref{sub-sec:event-window-identifier}. Based on these findings, we can state that Embed2Detect is suitable for real-time event detection. 


\section{Experimental study}
\label{sec:experiment-results}
In this section, we present the main results of the experiments which are conducted on social media data sets. More details about the data sets are described in Section \ref{sub-sec:data-sets-and-preparation}. To evaluate the results, the evaluation metrics mentioned under Section \ref{sub-sec:eval-metrics} were used. We implemented a prototype of Embed2Detect in Python 3.7 which has been made available on GitHub\footnote{Python implementation of Embed2Detect is available on \url{https://github.com/hhansi/embed2detect}}. 
Using this implementation, we analysed the performance of aggregation methods (Section \ref{sub-sec:aggregation-method}) and impact by text preprocessing (Section \ref{sub-sec:impact-preprocessing}). Furthermore, an analysis of parameter sensitivity was also conducted (Section \ref{sub-sec:param-selection}). To compare the performance of Embed2Detect with available methods, we considered three recent event detection methods from different competitive areas as baselines (Section \ref{sub-sec:baselines}). The corresponding comparison of results is reported under Section \ref{sub-sec:comparison-baselines}.  A comprehensive evaluation on the efficiency of Embed2Detect is available in Section \ref{sub-sec:efficiency-eval}. Finally, we conducted some experiments to suggest possible extensions to Embed2Detect using other word embedding models and the obtained results and suggestions are summarised in Section \ref{sub-sec:extensions-we}. All of the experiments were conducted on an Ubuntu 18.04 machine which has 2.40GHz 16 core CPU processor with 16GB RAM.

\vspace{8mm}
\subsection{Data sets and preparation}
\label{sub-sec:data-sets-and-preparation}
To conduct the experiments and evaluations, we used  data sets collected from Twitter. Considering some major issues associated with existing and available data sets, we decided to create and release our own data sets (Section \ref{sub-sec:data-sets}). Further details on data collection methods and data cleaning methods used in this research are mentioned in Section \ref{sub-sec:data-collection} and Section \ref{sub-sec:data-cleaning} respectively.

\subsubsection{Data sets} \label{sub-sec:data-sets}
The most recent data sets for social media event detection were released based on data in 2012 \citep{aiello2013sensing,mcminn2013building}. The data set released by \cite{mcminn2013building} (Events2012) is used the Twitter stream from 10 October 2012 to 7 November 2012. \cite{aiello2013sensing} released three data sets using filtered Twitter streams correspond to the domains of sports and politics. Due to the restrictions made by Twitter, both of these corpora only contained tweet IDs which can be used to download the tweets.

Downloading the tweets in Events2012 corpus, we could only retrieve 65.8\% of the tweets, as the rest were removed. In addition to the issue of missing a large proportion, a few more issues were encountered with this data set as follows. Since this data set was initially designed by considering event detection as identifying event clusters, only the event descriptions were provided as ground truth without temporal details. Therefore, following a commonly used strategy, we had to separate data into 1-day time windows\citep{alkhamees2016event, morabia2019sedtwik}. But, the 1-day time window is too lengthy to obtain quick updates targeted by our research. After time window separation, we found that all time windows are employed with events, due to the usage of a general data stream. The event occurred time window identification cannot be properly evaluated when all-time windows hold events.

Considering the data sets released by \cite{aiello2013sensing}, they were designed by focusing the same problem targeted by this research. But, similar to the Events2012 data set, a large proportion of data was found to be not available for downloading. For example, only 63.4\% of the sports data set could be downloaded. In addition to this issue, a major change to the tweet content was made in 2017, the character limit has been increased to 280 from 140. Thus, tweets in 2012, are comparatively small in character length compared to post 2017 tweets.

Considering the above-mentioned issues in available data sets, we decided to create new data sets to evaluate our approach. Also, we believed that releasing recent data sets would helpful to the research community too. Even though the proposed method is applicable to any social media data set, considering the restrictions, support and coverage given on data collection by different services, we decided to collect data from Twitter, similar to above-mentioned data sets. While generating the data sets, we focused on two different domains, namely, sports and politics to prove the domain independence of our method. Sports is known as a domain with rapid evolution and politics is known as a domain with a slow evolution \citep{adedoyin2016rule}. This domain selection is also motivated by the data sets released by \cite{aiello2013sensing}. More details on data collection are available in Section \ref{sub-sec:data-collection}. 

A similar strategy to \cite{aiello2013sensing} was used to ground truth (GT) preparation. We reviewed the published media reports related to the chosen topics during the targeted period and selected a set of events. Each event was supported using a set of keywords taken from news and social media to compare with the identified event words. We made these data sets including the GT labels publicly available\footnote{Data sets including the GT events are available on \url{https://github.com/hhansi/twitter-event-data-2019}}.

\subsubsection{Data collection} \label{sub-sec:data-collection}
Data collection was done using Twitter developer Application Programming Interfaces (APIs)\footnote{More details about Twitter developer service including its APIs are available at \url{https://developer.twitter.com/}}. Initially, data belonging to a particular topic was extracted using a trending hashtag. Then the hashtags found in the extracted data set were ranked based on their popularity and popular hashtags were used for further data extraction\footnote{For MUNLIV data collection, hashtags; \#MUNLIV, \#MUFC, \#LFC, \#Liverpool, \#GGMU, \#PL, \#VAR and \#YNWA were used and for BrexitVote data collection, hashtags; \#BrexitVote, \#SuperSaturday, \#Brexit, \#BrexitDeal, \#FinalSay, \#PeoplesVote, \#PeoplesVoteMarch were used.}.

To generate the sports data set, English Premier League 19/20 match between two popular teams, specifically, Manchester United and Liverpool was selected. This match was held at Old Trafford, Manchester on 20 October 2019. During the match, each team scored a single goal. Starting from 16:30, the total duration of the match was 115 minutes including the half time break. This data set will be referred to as `MUNLIV' in the following sections.  

To generate the political data set, Brexit Super Saturday in 2019 was selected. This was a UK parliament session which occurred on Saturday, 19 October 2019. It was the first Saturday session in 37 years. Even though it was organised to have a vote on a new Brexit deal, the vote was cancelled due to an amendment passed against the deal. This event started at 09:30 and held until around 16:30. This data set will be referred to as `BrexitVote' in the following sections.

For MUNLIV, we collected 118,700 tweets during the period 16:15-18:30. Among them, we used 99,995 (84.2\%) tweets posted during the match for experiments, because we could extract GT events only for this period using news media. For BrexitVote, we collected 276,448 tweets during the period 08:30-18:30 but only used 174,498 (63.1\%) tweets posted from the beginning of the parliament session until the vote on the amendment for experiments. Similar to the scenario with MUNLIV, the focus by news media was found to be high until the vote to extract more accurate GT events. Considering the evolution rate of each domain, for the sports data set MUNLIV, 2 minute, and for the political data set BrexitVote, 30 minute time windows are selected. After separating the data into chunks, on average there were 1,724 and 14,542 tweets per time window in sport and political data sets respectively. 

\subsubsection{Data cleaning} \label{sub-sec:data-cleaning}
To learn embeddings on separate tokens, embedding models need tokenised text. Since we focused on Twitter data sets during the experiments, we used the TweetTokenizer model available with Natural Language Toolkit (NLTK)\footnote{NLTK documentation is available at \url{https://www.nltk.org/}} to tokenise the text in tweets. This tokeniser was designed to be flexible on new domains with the consideration on characteristics in social media text such as repeating characters and special tokens. It has the ability to remove characters which repeats more than 3 times to generalise the various word forms introduced by users. For example, both words \textit{`goalll'} and \textit{`goallll'} will be replaced as \textit{`goalll'}. Further, it tokenises the emotions and words specific to social media context (e.g. 1-0, c'mon, \#LFC, :-)) correctly. Also, we did not preserve the case sensitivity in tokenised text.

In addition to tokenising, retweet notations, links and hash symbols were removed from the text. Retweet notations and links were removed because they do not make any contribution to the idea described. Hash symbols were removed to treat hashtags and other words similarly during embedding learning. To automate these removals, text pattern matching based on regular expressions was used.

\subsection{Evaluation metrics} \label{sub-sec:eval-metrics}
In order to evaluate the performance of the proposed method and baselines, event words are compared with GT event keywords using the following metrics. 
In the equations stated below, set of all event windows in the data set, detected event windows and relevant event windows found in detected windows are represented by $W$, $W^{d}$ and $W^{r}$ respectively.

\begin{itemize}
    \item \textbf{Recall}: Fraction of the number of relevant event windows detected among the total number of event windows that exist in the data set
    
    \begin{equation*}
        \text{Recall} = \frac{|W^{r}|}{|W|}
    \end{equation*}
    
    \item \textbf{Precision}: Fraction of the number of relevant event windows detected among the total number of event windows detected
    
    \begin{equation*}
        \text{Precision} = \frac{|W^{r}|}{|W^{d}|}
    \end{equation*}
    
    \item \textbf{F-Measure (F1)}: Weighted harmonic mean of precision and recall 
    
    \begin{equation*} \label{eg:f-measure}
        \text{F-Measure} = 2 \times \frac{precision \times recall}{precision + recall}
    \end{equation*}

    \item \textbf{Keyword Recall}: Fraction of the number of correctly identified words among the total number of keywords mentioned in the GT events \citep{aiello2013sensing}. To calculate a final value for a set of time windows, micro averaging \citep{manning2008text} is used. 
    
    \begin{equation*}
        \text{Keyword Recall} = \frac{\sum_{t \in T}|{w :w \in W_{t}^{d} \cap GT_{t}}|}{\sum_{t \in T} |w: w \in GT_{t}|}
    \end{equation*}
    
    $T$ represents the event occurred time windows, $w$ represents the words/ keywords and $GT$ represents the set of ground truth events.

\end{itemize}

While calculating the recall, precision and F-measure, a detected window is marked as a relevant event window, if all the events occurred during that time period are found in the event words identified for that window. A match between event words and a GT event is established if at least one GT keyword corresponding to that event is found from the event words. Likewise, for keyword recall calculation, if at least one word mentioned in a synonym (similar) word group in GT is found, it is considered as a match. Therefore, the total number of GT keywords is calculated as the total of synonym word groups.

\subsection{Aggregation method} \label{sub-sec:aggregation-method}
As mentioned in Section \ref{sub-sec:event-window-identifier}, to measure the temporal data change between two consecutive time windows, Embed2Detect needs to aggregate the values computed by cluster change calculation (Section \ref{sub-sec:cluster-change-calculation}) and vocabulary change calculation (Section \ref{sub-sec:vocabulary-change-calculation}). For this aggregation, we experimented the techniques: average and maximum considering their simplicity and common usage. The obtained results are shown in Table \ref{tab:results-aggregation}. 

\begin{table*}
\caption{Evaluation results with different aggregation methods}
\label{tab:results-aggregation}       
\begin{tabular}{l|ccc|ccc}
\hline\noalign{\smallskip}
\bf Data set & \multicolumn{3}{c|}{MUNLIV} & \multicolumn{3}{c}{BrexitVote} \\
\noalign{\smallskip}\hline\noalign{\smallskip}
\bf Method & \bf Recall & \bf Precision & \bf F1 & \bf Recall & \bf Precision & \bf F1 \\
\noalign{\smallskip}\hline\noalign{\smallskip}
average & 0.696 & 0.615 & 0.653 & 1.000 & 0.727 & 0.842 \\
maximum & 0.652 & 0.652 & 0.652 & 1.000 & 0.800 & 0.889 \\
\noalign{\smallskip}\hline
\end{tabular}
\end{table*}

According to the results, for MUNLIV data set, there is a slight change in F1 between average and maximum calculations. But, we can see balanced values for both recall and precision when the maximum is used. In BrexitVote, there is a clear change in F1, with higher value using the maximum calculation. Based on the observations made on two diverse domains, we decided to use maximum calculation as the default aggregation method in Embed2Detect.

\subsection{Text preprocessing}
\label{sub-sec:impact-preprocessing}
Even though preprocessing improves the effectiveness, it can strongly restrict the generality of a method. Therefore, we believe it is worth experimenting the impact of text preprocessing on Embed2Detect. To maintain the simplicity of our method, we only suggest two preprocessing techniques, removal of punctuation marks and stop words.  The evaluation results obtained with different configurations of these techniques are reported in Table \ref{tab:results-preprocess}. We only used cluster change calculation in these experiments, because it has a high influence by changes in tokens.

\begin{table*}
\caption{Evaluation results with different preprocessing techniques}
\label{tab:results-preprocess}       
\begin{tabular}{l|ccc|ccc}
\hline\noalign{\smallskip}
\bf Data set & \multicolumn{3}{c|}{MUNLIV} & \multicolumn{3}{c}{BrexitVote} \\
\noalign{\smallskip}\hline\noalign{\smallskip}
\bf Method & \bf Recall & \bf Precision & \bf F1 & \bf Recall & \bf Precision & \bf F1 \\
\noalign{\smallskip}\hline\noalign{\smallskip}
all tokens & 0.826 & 0.463 & 0.594 & 1.000 & 0.800 &  0.889 \\
without punctuation & 0.913 & 0.457 & 0.609 & 1.000 & 0.727 & 0.842 \\
\makecell[l]{without punctuation\\ and stop-words} & 0.696 & 0.552 &  0.615 & 1.000 & 0.800 & 0.889 \\
\noalign{\smallskip}\hline
\end{tabular}
\end{table*}

According to the obtained results, the highest F1 for both sport and political data sets is obtained by the tokens without punctuation and stop-words. Even though there is an improvement in the performance with preprocessing, these results show that we can obtain good measures without preprocessing also. This ability will be helpful in situations where we cannot integrate direct preprocessing mechanisms. For example, identifying events in low-resource language or multilingual data sets can be mentioned. However, to conduct the following experiments, we used the tokens without punctuation and stop-words, because both data sets used in this research are mainly written in English.

\subsection{Parameter sensitivity analysis} \label{sub-sec:param-selection}
In Embed2Detect, word embedding learner and event window identifier require some hyper-parameters. Sections \ref{sub-sec:param-we} and \ref{sub-sec:param-event-window} describe the impact of different hyper-parameter settings and heuristics behind their selections.

\subsubsection{Parameters - word embedding learning} \label{sub-sec:param-we}
Word embedding learning requires 3 hyper-parameters: minimum word count, context size, and vector dimension. Given a minimum word count, the learning phase ignores all tokens with less total frequency than the count. Context size defines the number of words around the word of interest to consider during the learning process. Vector dimension represents the number of neurons in the hidden layer which also will be used as the dimensionality of word embeddings.

\begin{figure}[ht]
\centering
\subfloat[F1 with different vector dimensions (with context size=5)]{
    \centering
    \includegraphics[width=0.46\textwidth]{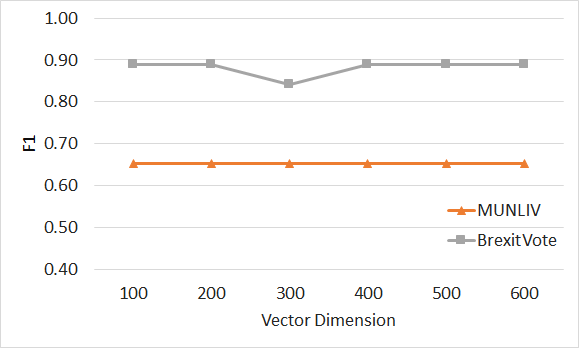}
    \label{sub-fig:f1-vector-dimension}
    }
\hfill    
\subfloat[F1 with different context sizes (with vector dimension=100)]{
    \centering
    \includegraphics[width=0.46\textwidth]{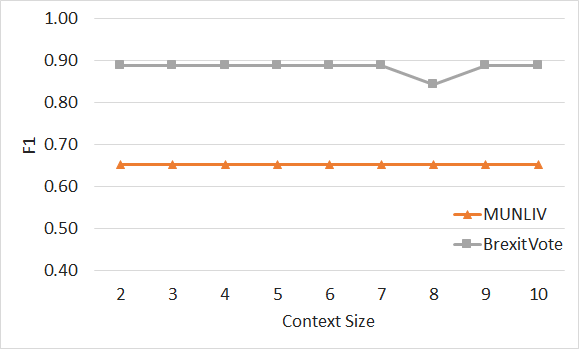}
    \label{sub-fig:f1-window-size}
    }

\subfloat[Time taken with different vector dimensions (with context size=5)]{
    \centering
    \includegraphics[width=0.46\textwidth]{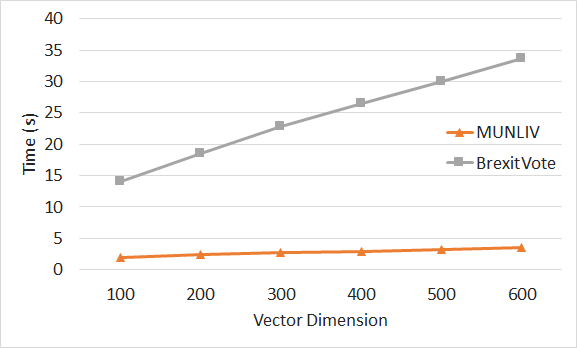}
    \label{sub-fig:time-vector-dimension}
    }
\hfill
\subfloat[Time taken with different context sizes (with vector dimension=100)]{
    \centering
    \includegraphics[width=0.46\textwidth]{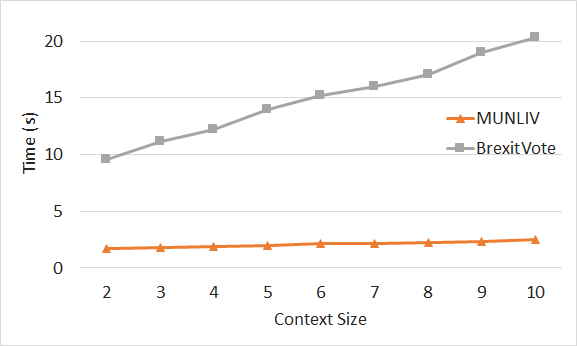}
    \label{sub-fig:time-window-size}
    }

\caption{Analysis on F1 and execution time with different values for word embedding learning parameters; vector dimension and context size (Average time taken to execute the full process on single data window is used for time values in both data sets)}

\label{fig:we-param-selection}
\end{figure}

Considering the limited amount of data available in a time window, we fixed the minimum word count to 1. Nevertheless, we analysed how the effectiveness and efficiency of event detection vary with different vector dimensions and context sizes to select optimal values. To evaluate the effectiveness, F-measure (F1) was used and results obtained for both data sets are visualised in Figure \ref{fig:we-param-selection}. Based on the results, there was no significant change in F1 with different vector dimensions and context sizes. But, there was a gradual increase in execution time when both hyper-parameter values are increased. 

Accuracy of text-similarity can be improved with the increase of both the amount of training data and vector dimensionality \citep{mikolov2013efficient}. Similarly, a larger context size can result in higher accuracy in text-similarity due to the provision of more training data \citep{li2017data}. But, these effects were not notably captured with event detection. As a major reason for this, we can mention that event detection is not that sensitive to the syntactical and semantical structure of text same as with text-similarity tasks. Also, since we train separate models for each time window, each model has comparatively small data sets to learn embedding space. Therefore, even though the vector dimensions are increased, no sufficient data will be provided for their proper adjustment.

Following the results we obtained and the findings of previous research \citep{mikolov2013efficient, li2017data}, we fixed 100 dimensions for word embeddings and length of 5 for context size. The decision on dimensionality is mainly influenced by the training data limitations and learning time. The context size is chosen considering the requirement of providing sufficient knowledge for learning and also the execution time. 

\subsubsection{Parameters - event window identification} \label{sub-sec:param-event-window}
As described in Section \ref{sub-sec:event-window-identifier}, the event window identifier requires two hyper-parameters, $\alpha$ and $\beta$. $\alpha$ is used to indicate the significance level of targeted events, and $\beta$ is used as a frequency threshold to remove the outlier tokens. Both of these hyper-parameters need to be set by the user or domain expert, considering the characteristics associated with the selected domain or filtered data stream.

\begin{figure*}
\centering
  \includegraphics[width=0.95\textwidth]{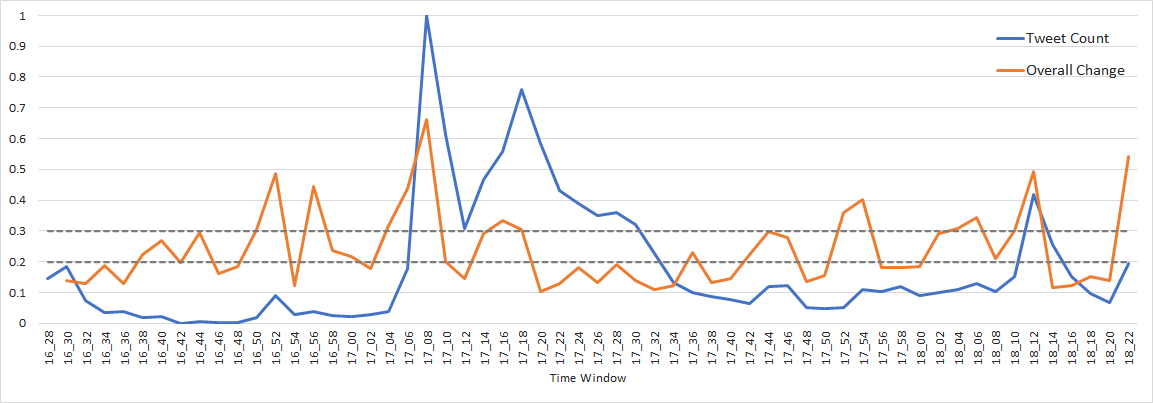}
  \caption{Overall change and tweet count variations over time windows - MUNLIV}
\label{fig:overall-change-time-munliv}       
\end{figure*}

\begin{figure*}
\centering
  \includegraphics[width=0.65\textwidth]{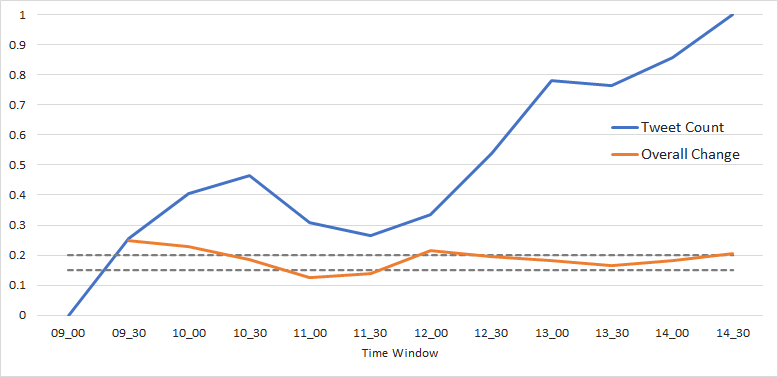}
  \caption{Overall change and tweet count variations over time windows - BrexitVote}
\label{fig:overall-change-time-brexitvote}       
\end{figure*}

The main idea behind event window identification is based on overall textual data change between time windows. A high overall change indicates the occurrence of a major event(s) and low change indicates minor event(s). To provide a clearer insight, we plotted the variations of temporal overall change values of data sets, MUNLIV and BrexitVote in Figure \ref{fig:overall-change-time-munliv} and \ref{fig:overall-change-time-brexitvote} respectively. Additionally, we plotted the total tweet count of each window in these graphs to highlight that the overall change-based measure is capable of identifying events which do not make a notable change to the total tweet count too. The total tweet count changes can only capture major events which make bursts. For comparison purpose, tweet counts are scaled down using the min-max normalisation. 

Focusing on Figure \ref{fig:overall-change-time-munliv} corresponds to MUNLIV data set, more fluctuations on overall change can be revealed, due to the rapid evolution in the sports domain. To do a deep analysis, at time window 16:40, a change of 0.269 and at 17:06, a change of 0.436 is measured. Looking at news media, at 16:40 a missed attempt and at 17:06 a goal is reported. Compared to the goal, a missed attempt is a minor event in the sports domain and overall change measure is capable of capturing that distinction successfully. Following this ability, $\alpha$ is used to filter the events based on user preference. For example, if $\alpha$ equals 0.2, both events missed attempt and goal will be captured by Embed2Detect. However, if the $\alpha$ value is increased to 0.3, among those two events, only the goal will be captured. There was no high number of fluctuations for BrexitVote data set (Figure \ref{fig:overall-change-time-brexitvote}) because the political domain has a comparatively slow evolution than the sports domain. Unlike with the MUNLIV data set, this effect limits the overall change values to a small range while increasing the sensitivity of $\alpha$. By slight variation of $\alpha$ (e.g. from 0.15 to 0.2), capturing events can be changed. 

Following these analyses, it is infeasible to define a common $\alpha$ value for different domains as well as for a particular domain. For different domains, this value needs to be picked, mainly considering the data evolution. Within a particular domain, $\alpha$ can be varied according to personal preferences on event importance. However, it can be simply chosen by using the domain knowledge and analysing a few past time windows.

\begin{figure*}
\centering
  \includegraphics[width=0.95\textwidth]{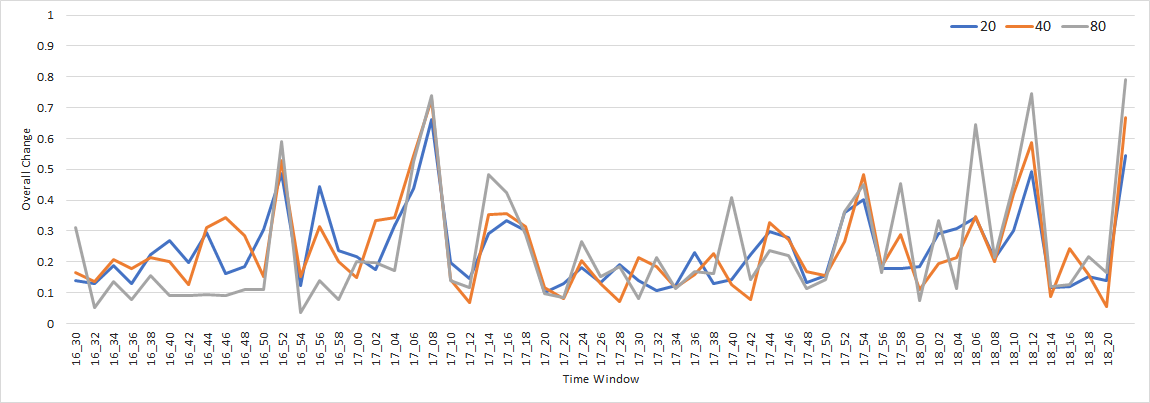}
  \caption{Analysis on impact by different $\beta$ values on overall temporal change}
\label{fig:overall-change-frequency-threshold}       
\end{figure*}

In addition to the $\alpha$ value, Embed2Detect uses another threshold $\beta$ to remove outlier tokens. Defined a $\beta$ value, all the tokens with less frequency that it, such as misspelt and uncommon words will be removed. The impact by different $\beta$ values on overall change measure is represented in Figure \ref{fig:overall-change-frequency-threshold}. According to this plot, with high $\beta$ values, some events can be missed (e.g. time window 16:40 will not be identified with $\beta=80$). Also, high $\beta$ could unnecessarily increase the overall change of some events (e.g. overall change increase at time window 18:06 and 18:12). 

\begin{figure*}
\centering
  \includegraphics[width=0.6\textwidth]{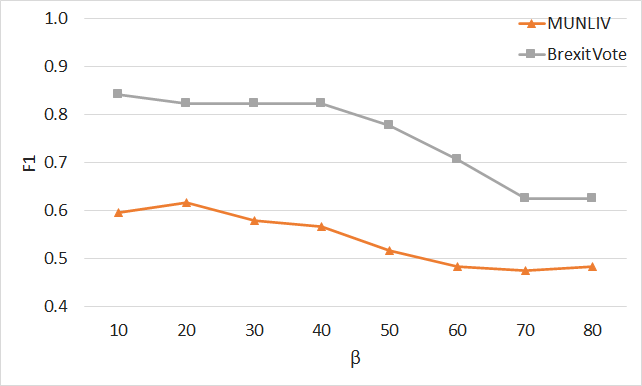}
  \caption{Analysis on F1 with different $\beta$ values (with $\alpha$=0.14)}
\label{fig:beta-f1}       
\end{figure*}

The major reason behind these behaviours is the removal of event tokens with high $\beta$ value. Due to this, the effectiveness of event detection decreases with increasing $\beta$ (Figure \ref{fig:beta-f1}). Therefore, $\beta$ value only need to be sufficiently large to remove outliers. Analysing the sport and political data sets, values less than 20 is appropriate for $\beta$. But similar to $\alpha$, $\beta$ also highly depend on domain-specific characteristics such as word usage and audience. Therefore, we believe that $\beta$ is also a hyper-parameter which needs to be controlled by domain experts. 

\subsection{Baseline methods} \label{sub-sec:baselines}
Since there is no specific data set to evaluate event detection performance, available methods cannot be compared with each other to pick the best baseline. Therefore, considering the requirements of event detection and available competitive areas, we selected three recently proposed methods as baselines. The major requirements we focused on during this selection were effectiveness, efficiency and expandability. We also covered different competitive areas which can be summarised as the incorporation of the social aspect, word acceleration over frequency, unsupervised learning (tensor decomposition and clustering) and segments over uni-grams, to make the baselines strong enough. All of these methods process the whole data stream without considering only some keywords (e.g. hashtags) to identify temporal events, similar to our approach. More details on selected baseline methods are as follows.

\begin{itemize}
    \item \textbf{MABED}
    \citep{guille2015event}: Anomalous user mention-based statistical method \\
    Mention anomalies were taken into consideration in this research in order to incorporate the social aspect of Twitter with event detection rather than only focusing on textual contents of tweets. User mentions are links added intentionally to connect a user with a discussion or dynamically during re-tweeting. Anomalous variations in mention creation frequency and their magnitudes were used for event detection. To extract the event words, co-occurrences of words and their temporal dynamics were used.
    
    \item \textbf{TopicSketch}  \citep{xie2016topicsketch}: Word acceleration-based tensor decomposition method\\
    Word acceleration is suggested by this research to support event detection because it has the ability to differentiate bursty topics (events) from general topics like car, food, or music. Events have the ability to force people to discuss them intensively. This force can be expressed by acceleration and this research proposed it as a good measure over frequency for event detection. To extract the event words, a tensor decomposition method, SVD was used. 
    
    \item \textbf{SEDTWik} \citep{morabia2019sedtwik}: Segment-based clustering method powered by Wikipedia page titles \\
    This is an extension to the Twevent system \citep{li2012twevent}. Text segments are focused in this research because they are more meaningful and specific than uni-grams. Wikipedia page titles were used as a semantic resource during segment extraction to preserve the informativeness of identified segments. To identify the events, bursty segments were clustered using Jarvis-Patrick algorithm. Burstiness of segments is calculated using both text statistics and user diversity-based measures. 
\end{itemize}

\subsection{Comparison with baselines} \label{sub-sec:comparison-baselines}
We compared the effectiveness and efficiency of Embed2Detect with selected baseline methods: MABED, TopicSketch and SEDTWik (Section \ref{sub-sec:baselines}). Effectiveness was measured using the evaluation metrics: recall, precision, F1 and keyword recall (Section \ref{sub-sec:eval-metrics}). To measure the efficiency, total time taken to execute the complete process on full data sets and the average time taken per time window by each method were used.

Similar to the hyper-parameters $\alpha$ and $\beta$ in Embed2Detect, all the baseline methods have their own parameters which need to be optimised depending on the data set. Therefore, to generate comparable results, a common strategy is used to identify optimal hyper-parameters, because they make a high impact on the method's performance. For each method, we evaluated all possible hyper-parameter settings to choose the best F1 value. For MABED, we optimised the hyper-parameters: number of events ($k$), maximum number of words describing each event ($p$), weight threshold for selecting relevant words ($\theta$) and overlap threshold ($\sigma$). For $k$ and $p$, starting from a low value we kept increasing them gradually until the maximum F1, which reduces with further increasing parameter values is reached. Similarly, for $\theta$ and $\sigma$, we experimented the values around the original values reported in initial experiments \citep{guille2015event}. For TopicSketch, we decided to optimise only the most critical hyper-parameter due to the high time complexity of this method. Thus, while keeping default values for other parameters, we tested gradually increasing values for detection threshold to obtain the best F1 value. For SEDTWik, we optimised the hyper-parameters: number of subwindows ($M$), number of cluster neighbours ($k$) and newsworthiness threshold ($\tau$). Different values for $M$ are picked considering the time windows lengths assigned to each data sets. For $k$ and $\tau$, starting from a low value, gradually increasing values were tested to obtain highest F1. For Embed2Detect, we identified all possible values for each hyper-parameter $\alpha$ and $\beta$, and experimented with all of their combinations to get best results. Following these parameter optimisations, results obtained for MUNLIV and BrexitVote are reported in Tables \ref{tab:results-munliv} and \ref{tab:results-brexitvote} respectively. The corresponding parameter settings are summarised in Table \ref{tab:param-setting}. To measure the reported execution times, we used sequential processing for the baseline methods according to  the available implementations and parallel processing with 8 workers for Embed2Detect.  Both total time taken to process the whole data stream and average time taken per time window are reported. 

\begin{table*}
\caption{Performance comparison of Embed2Detect with baseline methods using MUNLIV data set}
\label{tab:results-munliv}       
\begin{tabular}{l|cccc|cc}
\hline\noalign{\smallskip}
\bf \multirow{2}{*}{Method} & \bf \multirow{2}{*}{Recall} & \bf \multirow{2}{*}{Precision} & \bf \multirow{2}{*}{F1} & \bf \makecell[l]{Keyword\\ Recall} & \multicolumn{2}{c}{\bf Execution Time(s)} \\ 
& & & & & Total & Average \\
\noalign{\smallskip}\hline\noalign{\smallskip}
MABED & 0.478 & 0.193 & 0.275 & 0.348 & \bf 168 & \bf 2.947 \\
TopicSketch & 0.609 & 0.246 & 0.350 & 0.400 & 25492 & 447.228 \\
SEDTWik & 0.652 & 0.268 & 0.380 & 0.386 & 1290 & 22.632 \\
Embed2Detect & \bf 0.652 & \bf 0.652 & \bf 0.652 & \bf 0.843 & 202 & 3.544 \\
\noalign{\smallskip}\hline
\end{tabular}
\end{table*}

\begin{table*}
\caption{Performance comparison of Embed2Detect with baseline methods using BrexitVote data set}
\label{tab:results-brexitvote}       
\begin{tabular}{l|cccc|cc}
\hline\noalign{\smallskip}
\bf \multirow{2}{*}{Method} & \bf \multirow{2}{*}{Recall} & \bf \multirow{2}{*}{Precision} & \bf \multirow{2}{*}{F1} & \bf \makecell[l]{Keyword\\ Recall} & \multicolumn{2}{c}{\bf Execution Time(s)} \\ 
& & & & & Total & Average \\
\noalign{\smallskip}\hline\noalign{\smallskip}
MABED & 0.625 & 0.455 & 0.526 & 0.403 & 532 & 48.364 \\
TopicSketch & 0.500 & 0.364 & 0.421 & 0.254 & 15887& 1444.273 \\
SEDTWik & 0.750 & 0.500 & 0.600 & 0.426 & 702 & 63.818 \\
Embed2Detect & \bf 1.000 & \bf 0.800 & \bf 0.889 & \bf 0.985 & \bf 310 &\bf 28.182 \\
\noalign{\smallskip}\hline
\end{tabular}
\end{table*}

\begin{table*}
\caption{Parameter settings used by each method for the best results}
\label{tab:param-setting}       
\begin{tabular}{l|l|l}
\hline\noalign{\smallskip}
\bf Method & \bf \makecell[l]{Parameter Setting\\ (MUNLIV)} & \bf \makecell[l]{Parameter Setting\\ (BrexitVote)} \\
\noalign{\smallskip}\hline\noalign{\smallskip}
MABED & \makecell[l]{k = 150\\ p = 20\\ $\theta$ = 0.7\\ $\sigma$ = 0.5} & \makecell[l]{k = 150\\ p = 20\\ $\theta$ = 0.6\\ $\sigma$ = 0.5}  \\
\hline
TopicSketch & \makecell[l]{detection threshold = 60\\ bucket size = 5000} & \makecell[l]{detection threshold = 35\\ bucket size = 5000} \\ 
\hline
SEDTWik & \makecell[l]{M = 2\\ k = 6\\ $\tau$ = 0.7\\} & \makecell[l]{M = 2\\ k = 6\\ $\tau$ = 0.2\\}\\
\hline
Embed2Detect & \makecell[l]{$\beta$ = 20\\ $\alpha$ = 0.23} & \makecell[l]{$\beta$ = 10\\ $\alpha$ = 0.16}  \\
\noalign{\smallskip}\hline
\end{tabular}
\end{table*}

Embed2Detect outperforms the baseline methods in both data sets with F1 of 0.652 on MUNLIV and F1 of 0.889 on BrexitVote. This proves that our method has the ability to detect the events effectively in diverse domains, specifically, sports and politics than the available methods. According to the recall and precision measures, all methods tends to return high recall than precision. When preparing the GT events based on news reports, there is a possibility to miss some important events which are only discussed within the social media platform \citep{aiello2013sensing, morabia2019sedtwik}. Due to that, some actual events can be labelled as false positives and it will reduce the precision value. Considering recall, the majority of the methods (except TopicSketch) resulted in high values with BrexitVote than MUNLIV data set. Comparing the GT of two data sets, BrexitVote has 72.7\% of event occurred time windows while MUNLIV has only 40.4\%. Due to this bias, high recall can be resulted with BrexitVote data set. Theoretically, such bias is captured because of the low evolution and less dynamicity in the political domain.

Following the execution times, for MUNLIV, MABED took 168 seconds and Embed2Detect took 34 seconds more than MABED. But on BrexitVote, Embed2Detect completed the execution in 310 seconds -- 222 seconds faster than MABED. In terms of average execution time per window, Embed2Detect took 2.947 seconds to process a 2-minute window in MUNLIV data set and 28.182 seconds to process a 30-minute window in BrexitVote data set. These time measures prove that Embed2Detect is sufficiently fast for real-time event detection. More detailed analysis of intermediate processing time of Embed2Detect is reported in Appendix \ref{appendix-intermidiate-time}.

\subsection{Efficiency evaluation} \label{sub-sec:efficiency-eval}
\begin{figure*}
\centering
  \includegraphics[width=0.6\textwidth]{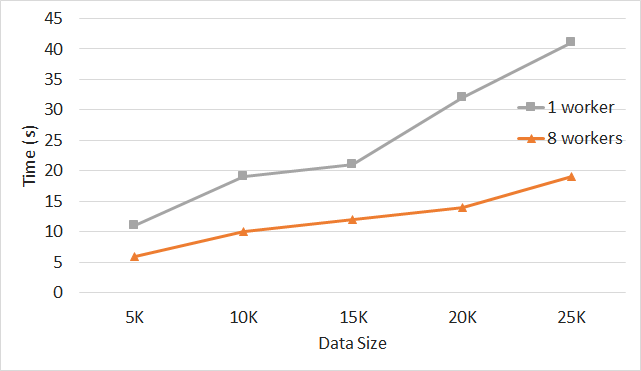}
  \caption{Execution time on different data sizes including the effect by sequential and parallel processing}
\label{fig:data-time}       
\end{figure*}

Processing time is a critical measure in real-time applications. For successful event detection, the huge amount of data generated in social media needs to be processed in (near) real-time. Considering the problem targeted by Embed2Detect, efficiency requirement can be further specified as processing the data belong to a time window within a sufficiently short period. Following this requirement, we evaluated the scalability of Embed2Detect and a parallelised version of Embed2Detect by measuring their execution times for the complete process on time windows with increasing data size. The obtained results are plotted in Figure \ref{fig:data-time}. As the data size within a time window (e.g. 1-minute window), 5,000 to 25,000 tweets were considered. Focusing on a filtered data stream, the upper limit of 25,000 tweets can be mentioned as a reasonable amount to depict the real scenario.

According to the results, the sequential version of Embed2Detect took nearly 10 seconds to process 5,000 tweets and this increased to 41 seconds to process 25,000 tweets. The parallel version with eight workers reduced the processing time to 6 seconds for 5,000 tweets and 19 seconds for 25,000 tweets. Also, we noticed that for both implementations, sequential and parallel, execution time grew linearly with data size (Figure \ref{fig:data-time}). Following these results, we can confirm that our approach is adequately efficient to facilitate real-time processing. Due to the linear growth of execution time, we can guarantee that Embed2Detect is capable of handling data bursts too.

\subsection{Extension to other word embedding models} \label{sub-sec:extensions-we}
Word embedding models other than Skip-gram can also be used with Embed2Detect. Since we implemented word embedding learner as a separate module in the Embed2Detect architecture, different word embeddings can be easily connected. But, it is important to consider the learning time and associated complexities while selecting a word embedding model to satisfy the goal of real-time event detection. In this section, we discuss the appropriateness of different architectures for word embedding generation in Embed2Detect. 

For this analysis, we used fastText, BERT and DistilBERT models. FastText is an updated version of the Skip-gram model which considers subword information while learning word representations \citep{bojanowski2017enriching}. Both BERT and DistilBERT are transformer-based models. According to the recent advances in the domain of NLP, transformers gained success in many areas such as language generation \citep{devlin-etal-2019-bert}, named entity recognition \citep{liang2020bond} and question answering \citep{yang-etal-2019-end}. BERT: Bidirectional Encoder Representations from Transformers \citep{devlin-etal-2019-bert} is the first transformer model which gained wide attention. This model is designed to train from unlabelled text using the masked language modelling (MLM) objective and to fine-tune for a downstream task, as a solution for the high data requirement by deep neural networks. DistilBERT is a distilled version of BERT which is light and fast \citep{sanh2019distilbert}.

\begin{table*}
\caption{Time taken to learn embeddings by different architectures}
\label{tab:embedding-learning-time}       
\begin{tabular}{l|l|cccc}
\hline\noalign{\smallskip}
\multirow{2}{*}{\makecell[l]{\bf Time Window\\ \bf Length}} & \multirow{2}{*}{\bf Tweet Count} & \multicolumn{4}{c}{\bf Embedding Learning Time (s)}\\
&  & Skip-gram & fastText & BERT & DistilBERT \\
\noalign{\smallskip}\hline\noalign{\smallskip}
2 min.(120 s) & \makecell[c]{1705} & 1 & 12 & 646 & 433\\
30 min.(1800 s) & \makecell[c]{20133} & 18 & 41 & 21442 & 11699\\
\noalign{\smallskip}\hline
\end{tabular}
\end{table*}

Initially, the time taken by different architectures to learn word embeddings is measured and obtained results are summarised in Table \ref{tab:embedding-learning-time}. Both Skip-gram and fastText models were trained from scratch using Twitter data as suggested by this research. Following the idea presented with transformers, for both BERT and DistilBERT, we retrained available models using our data. As the pre-trained BERT model, \textit{bert-base-uncased} and DistilBERT model, \textit{distilbert-base-uncased} released by HuggingFace’s Transformers library \citep{wolf2019huggingface} are selected. According to the obtained results, classic word embedding models (e.g. Skip-gram and fastText) learn the representations faster than transformer-based models (e.g. BERT and DistilBERT).

Comparing fastText and Skip-gram, fastText took more time because it processes subword information. But, incorporation of subwords allows this model to capture connections between modified words. For example, consider the goal-related words found within the top 20 words with high cluster change during a goal score:

\begin{quote}
    Skip-gram- \textit{goal, goalll, rashyyy, scores} \\
    fastText- \textit{goalll, goooaaalll, rashford, rashyyy, @marcusrashford, scored, scores}
\end{quote}

\noindent fastText captures more modified words than Skip-gram. We could not run a complete evaluation using fastText embeddings, because it requires a manual process since GT keywords only contain the words in actual form. 

Transformer-based models took more time than both Skip-gram and fastText due to their complex architecture to learn contextualised word embeddings. DistilBERT is found to be faster than BERT, however, the learning time of DistilBERT is not fast enough for real-time processing because it exceeds the tweet generation time. For example to learn from tweets posted during a 2-minute time window, it took approximately 7.2 minutes. If this model can be further distilled, there is a possibility to achieve the required efficiency to become suitable for real-time processing. However, further distillation can reduce the language understanding capability of the model as there is a 3\% reduction in DistilBERT compared to BERT \citep{sanh2019distilbert}.

According to recent literature, transformer-based models performed well on many  NLP-related tasks, because of the ability to capture the contextual sense of words. BERT is capable of generating different embeddings for the same word depending on its surrounding context. In other words, the main idea behind BERT is capturing spacial changes of words. From the perspective of processing formally written natural language, this is a very useful feature. But, in social media, language is mostly informal and for event detection using social media text, temporal changes of words need to be more focused. If we consider a particular time window of a filtered data stream, it is rarely possible to have a word with two totally different contextual meanings. Therefore, the context awareness associated with BERT is not much useful for event detection. 

\begin{figure*}[ht]
\centering
  \includegraphics[width=0.95\textwidth]{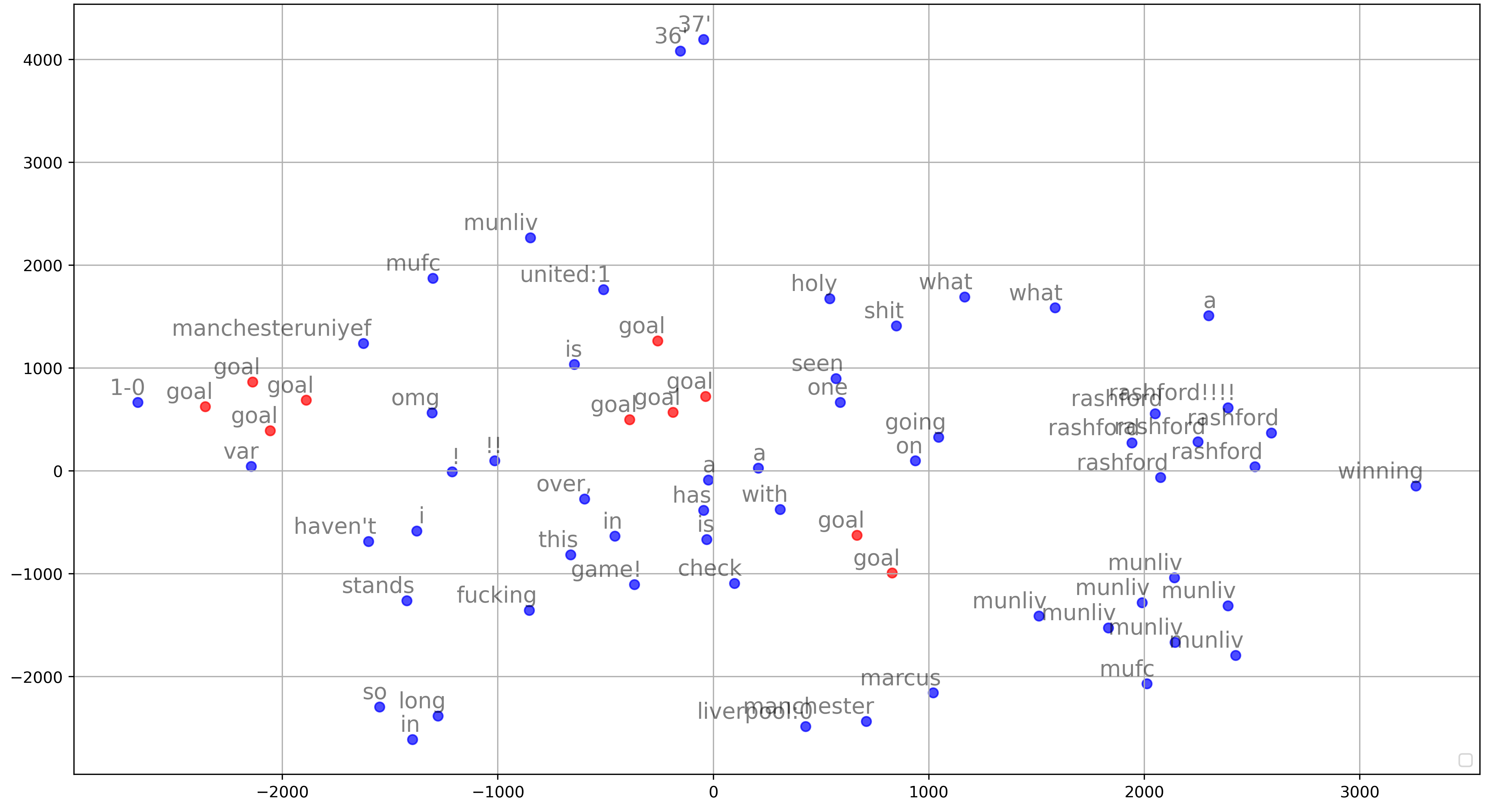}
  \caption{t-SNE visualisation of sample word embeddings obtained by a \textit{bert-base-uncased} model which is retrained on MUNLIV goal scored time window 2019-10-20 17:06 - 17:08}
\label{fig:bert-word-visualisation}       
\end{figure*}

Further, contextualised word embeddings could incorporate an additional complexity to the event detection method. For example, during a goal scoring of a football match, the word \textit{`goal'} will be expressed by the audience of the winning team and losing team differently. Even though the surrounding contexts are varied, the meaning of the word \textit{`goal'} targeted by event detection is constant. For such a scenario, BERT will return different embeddings for \textit{`goal'} as illustrated in Figure \ref{fig:bert-word-visualisation}. Having multiple embeddings for monosemy words can confuse the clusters and increase the computational complexity of the method exponentially. To overcome these issues, multiple embeddings of a word can be combined using an aggregation method. But, it breaks the main objective of contextualised word embeddings. Therefore, we believe it is inaccurate to aggregate context-aware embeddings. Following these findings, we can conclude that contextualised word embedding models such as BERT are less appropriate to be used with Embed2Detect, due to their
complexities which are not necessary for event detection. 

\section{Conclusions and future work}
\label{sec:conclusion}
In this paper, we proposed a novel event detection method coined Embed2Detect to identify the events occurred in social media data streams. Embed2Detect mainly combines the characteristics in word embeddings and hierarchical agglomerative clustering. This method uses self-learned word embeddings to capture the features in the targeted corpus in order to facilitate domain, platform or language-independent event detection.  Therefore, Embed2Detect can be easily applied on any social media data set in any language even though the majority of available methods are limited to specific platforms (e.g. Twitter) and languages (e.g. English). Further, this approach is also applicable to multilingual data sets. The ability to process multilingual data sets can be highlighted as an important requirement to process the data in social media considering its user base which is distributed all over the world. 

In contrast with prior work, Embed2Detect not only considers syntax and statistics in the underlying text but also incorporates semantics. Inclusion of semantics allows to understand the relationships between words. Due to the huge and diverse user base, social media text contains different words and word sequences which describe the same idea. Knowing the relationships between words, differently described similar ideas and their connections can be extracted. Therefore, our approach is capable to reduce the information loss experienced in previous approaches due to the lack of semantic involvement. 

According to the evaluations conducted, Embed2Detect performed significantly better than the recently suggested event detection methods, namely, MABED, TopicSketch and SEDTWik on both data sets MUNLIV and BrexitVote from the domain of sports and politics. As evaluation metrics, we used recall, precision F-measure and keyword recall to conduct a comprehensive evaluation. Also, we considered data from two contrasting domains which have different word usage, audience and evolution rate to evaluate the universality of methods. In addition to focusing on effectiveness, we measured the efficiency of Embed2Detect also, because real-time event detection is a time-critical operation. Embed2Detect performed event detection in both data sets within a short time period and it could handle increasing data volume to indicate its appropriateness for real-time application. In summary, the results we obtained from the experiments conclude that Embed2Detect can detect the events in social media data effectively and efficiently without depending on domain-specific features.

As an extension to Embed2Detect, more advanced word embedding learning methods can be applied. But, considering the learning time and associated complexities, to preserve the efficiency of the method, it is more suitable to use classic word embedding models such as Skip-gram than advanced word embedding models such as BERT. Under classic word embeddings, we hope to further analyse the impact by subword and character-based models which can be used to capture the connections between informal or modified text and their formal versions. Such an approach would be useful to understand informal text which is common to the context of social media. Further, focusing on the recent improvements to the domain of NLP by available transformer-based models, their pre-trained word embeddings can be supported to generate more comprehensive event details such as summaries using the detected event words in a future phase of this research. Also, we plan to further extend our method to identify event evolution over time to facilitate both event detection and tracking together.

\section*{Compliance with ethical standards}
\textbf{Conflict of Interest:} The authors declare that they have no conflict of interest.


\bibliographystyle{spbasic-ml}
\bibliography{references.bib}

\begin{thebibliography}{62}
\providecommand{\natexlab}[1]{#1}
\providecommand{\url}[1]{{#1}}
\providecommand{\urlprefix}{URL }
\expandafter\ifx\csname urlstyle\endcsname\relax
  \providecommand{\doi}[1]{DOI~\discretionary{}{}{}#1}\else
  \providecommand{\doi}{DOI~\discretionary{}{}{}\begingroup
  \urlstyle{rm}\Url}\fi
\providecommand{\eprint}[2][]{\url{#2}}

\bibitem[{Adedoyin-Olowe et~al(2013)Adedoyin-Olowe, Gaber, and
  Stahl}]{adedoyin2013trcm}
Adedoyin-Olowe M, Gaber MM, Stahl F (2013) Trcm: a methodology for temporal
  analysis of evolving concepts in twitter. In: International Conference on
  Artificial Intelligence and Soft Computing, Springer, pp 135--145

\bibitem[{Adedoyin-Olowe et~al(2016)Adedoyin-Olowe, Gaber, Dancausa, Stahl, and
  Gomes}]{adedoyin2016rule}
Adedoyin-Olowe M, Gaber MM, Dancausa CM, Stahl F, Gomes JB (2016) A rule
  dynamics approach to event detection in twitter with its application to
  sports and politics. \emph{Expert Systems with Applications} 55:351--360

\bibitem[{Aiello et~al(2013)Aiello, Petkos, Martin, Corney, Papadopoulos,
  Skraba, G{\"o}ker, Kompatsiaris, and Jaimes}]{aiello2013sensing}
Aiello LM, Petkos G, Martin C, Corney D, Papadopoulos S, Skraba R, G{\"o}ker A,
  Kompatsiaris I, Jaimes A (2013) Sensing trending topics in twitter.
  \emph{IEEE Transactions on Multimedia} 15(6):1268--1282

\bibitem[{Aldhaheri and Lee(2017)}]{aldhaheri2017event}
Aldhaheri A, Lee J (2017) Event detection on large social media using temporal
  analysis. In: 2017 IEEE 7th Annual Computing and Communication Workshop and
  Conference (CCWC), IEEE, pp 1--6

\bibitem[{Alkhamees and Fasli(2016)}]{alkhamees2016event}
Alkhamees N, Fasli M (2016) Event detection from social network streams using
  frequent pattern mining with dynamic support values. In: 2016 IEEE
  International Conference on Big Data (Big Data), IEEE, pp 1670--1679

\bibitem[{Antoniak and Mimno(2018)}]{antoniak2018evaluating}
Antoniak M, Mimno D (2018) Evaluating the stability of embedding-based word
  similarities. \emph{Transactions of the Association for Computational
  Linguistics} 6:107--119

\bibitem[{Bengio et~al(2003)Bengio, Ducharme, Vincent, and
  Jauvin}]{bengio2003neural}
Bengio Y, Ducharme R, Vincent P, Jauvin C (2003) A neural probabilistic
  language model. \emph{Journal of machine learning research} 3(Feb):1137--1155

\bibitem[{Benhardus and Kalita(2013)}]{benhardus2013streaming}
Benhardus J, Kalita J (2013) Streaming trend detection in twitter.
  \emph{International Journal of Web Based Communities} 9(1):122--139

\bibitem[{Bojanowski et~al(2017)Bojanowski, Grave, Joulin, and
  Mikolov}]{bojanowski2017enriching}
Bojanowski P, Grave E, Joulin A, Mikolov T (2017) Enriching word vectors with
  subword information. \emph{Transactions of the Association for Computational
  Linguistics} 5:135--146

\bibitem[{Castillo et~al(2011)Castillo, Mendoza, and
  Poblete}]{castillo2011information}
Castillo C, Mendoza M, Poblete B (2011) Information credibility on twitter. In:
  Proceedings of the 20th international conference on World wide web, ACM, pp
  675--684

\bibitem[{Chaffey(2019)}]{dave2019global}
Chaffey D (2019) Global social media research summary 2019 | smart insights.
  \url{https://www.smartinsights.com/social-media-marketing/social-media-strategy/new-global-social-media-research/}

\bibitem[{Chen et~al(2017)Chen, Kong, and Mao}]{chen2017online}
Chen G, Kong Q, Mao W (2017) Online event detection and tracking in social
  media based on neural similarity metric learning. In: 2017 IEEE International
  Conference on Intelligence and Security Informatics (ISI), IEEE, pp 182--184

\bibitem[{Choi and Park(2019)}]{choi2019emerging}
Choi HJ, Park CH (2019) Emerging topic detection in twitter stream based on
  high utility pattern mining. \emph{Expert Systems with Applications}
  115:27--36

\bibitem[{Clement(2019)}]{clement2019global}
Clement J (2019) Global social media ranking 2019 | statista.
  \url{https://www.statista.com/statistics/272014/global-social-networks-ranked-by-number-of-users/}

\bibitem[{Comito et~al(2019{\natexlab{a}})Comito, Forestiero, and
  Pizzuti}]{comito2019bursty}
Comito C, Forestiero A, Pizzuti C (2019{\natexlab{a}}) Bursty event detection
  in twitter streams. \emph{ACM Transactions on Knowledge Discovery from Data
  (TKDD)} 13(4):1--28

\bibitem[{Comito et~al(2019{\natexlab{b}})Comito, Forestiero, and
  Pizzuti}]{comito2019word}
Comito C, Forestiero A, Pizzuti C (2019{\natexlab{b}}) Word embedding based
  clustering to detect topics in social media. In: 2019 IEEE/WIC/ACM
  International Conference on Web Intelligence (WI), IEEE, pp 192--199

\bibitem[{Corney et~al(2014)Corney, Martin, and G{\"o}ker}]{corney2014spot}
Corney D, Martin C, G{\"o}ker A (2014) Spot the ball: Detecting sports events
  on twitter. In: European Conference on Information Retrieval, Springer, pp
  449--454

\bibitem[{Devlin et~al(2019)Devlin, Chang, Lee, and
  Toutanova}]{devlin-etal-2019-bert}
Devlin J, Chang MW, Lee K, Toutanova K (2019) {BERT}: Pre-training of deep
  bidirectional transformers for language understanding. In: Proceedings of the
  2019 Conference of the North {A}merican Chapter of the Association for
  Computational Linguistics: Human Language Technologies, Volume 1 (Long and
  Short Papers), Association for Computational Linguistics, Minneapolis,
  Minnesota, pp 4171--4186, \doi{10.18653/v1/N19-1423},
  \urlprefix\url{https://www.aclweb.org/anthology/N19-1423}

\bibitem[{Edouard et~al(2017)Edouard, Cabrio, Tonelli, and
  Le-Thanh}]{edouard2017graph}
Edouard A, Cabrio E, Tonelli S, Le-Thanh N (2017) Graph-based event extraction
  from twitter. In: Proceedings of the International Conference Recent Advances
  in Natural Language Processing, RANLP 2017, pp 222--230

\bibitem[{Godin et~al(2015)Godin, Vandersmissen, De~Neve, and Van~de
  Walle}]{godin2015multimedia}
Godin F, Vandersmissen B, De~Neve W, Van~de Walle R (2015) Multimedia lab@ acl
  wnut ner shared task: Named entity recognition for twitter microposts using
  distributed word representations. In: Proceedings of the workshop on noisy
  user-generated text, pp 146--153

\bibitem[{Gottfried and Shearer(2017)}]{Gottfried2017NewsUA}
Gottfried JA, Shearer E (2017) News use across social media platforms 2017.
  \url{https://www.journalism.org/2017/09/07/news-use-across-social-media-platforms-2017/}

\bibitem[{Guille and Favre(2015)}]{guille2015event}
Guille A, Favre C (2015) Event detection, tracking, and visualization in
  twitter: a mention-anomaly-based approach. \emph{Social Network Analysis and
  Mining} 5(1):18

\bibitem[{Hasan et~al(2018)Hasan, Orgun, and Schwitter}]{hasan2018survey}
Hasan M, Orgun MA, Schwitter R (2018) A survey on real-time event detection
  from the twitter data stream. \emph{Journal of Information Science}
  44(4):443--463

\bibitem[{Hasan et~al(2019)Hasan, Orgun, and Schwitter}]{hasan2019real}
Hasan M, Orgun MA, Schwitter R (2019) Real-time event detection from the
  twitter data stream using the twitternews+ framework. \emph{Information
  Processing \& Management} 56(3):1146--1165

\bibitem[{James(2019)}]{james2019data}
James J (2019) Data never sleeps 7.0. 2019.
  \url{https://www.domo.com/learn/data-never-sleeps-7}

\bibitem[{Kwak et~al(2010)Kwak, Lee, Park, and Moon}]{kwak2010twitter}
Kwak H, Lee C, Park H, Moon S (2010) What is twitter, a social network or a
  news media? In: Proceedings of the 19th international conference on World
  wide web, AcM, pp 591--600

\bibitem[{Li et~al(2012)Li, Sun, and Datta}]{li2012twevent}
Li C, Sun A, Datta A (2012) Twevent: segment-based event detection from tweets.
  In: Proceedings of the 21st ACM international conference on Information and
  knowledge management, pp 155--164

\bibitem[{Li et~al(2014)Li, Tai, Zhang, Yu, and Liu}]{li2014online}
Li J, Tai Z, Zhang R, Yu W, Liu L (2014) Online bursty event detection from
  microblog. In: 2014 IEEE/ACM 7th International Conference on Utility and
  Cloud Computing, IEEE, pp 865--870

\bibitem[{Li et~al(2017{\natexlab{a}})Li, Nourbakhsh, Shah, and
  Liu}]{li2017real}
Li Q, Nourbakhsh A, Shah S, Liu X (2017{\natexlab{a}}) Real-time novel event
  detection from social media. In: 2017 IEEE 33rd International Conference on
  Data Engineering (ICDE), IEEE, pp 1129--1139

\bibitem[{Li et~al(2017{\natexlab{b}})Li, Shah, Liu, and
  Nourbakhsh}]{li2017data}
Li Q, Shah S, Liu X, Nourbakhsh A (2017{\natexlab{b}}) Data sets: Word
  embeddings learned from tweets and general data. \emph{arXiv preprint
  arXiv:170803994}

\bibitem[{Liang et~al(2020)Liang, Yu, Jiang, Er, Wang, Zhao, and
  Zhang}]{liang2020bond}
Liang C, Yu Y, Jiang H, Er S, Wang R, Zhao T, Zhang C (2020) Bond:
  Bert-assisted open-domain named entity recognition with distant supervision.
  In: Proceedings of the 26th ACM SIGKDD International Conference on Knowledge
  Discovery \& Data Mining, pp 1054--1064

\bibitem[{Maaten and Hinton(2008)}]{maaten2008visualizing}
Maaten Lvd, Hinton G (2008) Visualizing data using t-sne. \emph{Journal of
  machine learning research} 9(Nov):2579--2605

\bibitem[{Manning et~al(2008{\natexlab{a}})Manning, Raghavan, and
  Sch{\"u}tze}]{manning2008introduction}
Manning CD, Raghavan P, Sch{\"u}tze H (2008{\natexlab{a}}) \emph{Introduction
  to information retrieval}. Cambridge university press

\bibitem[{Manning et~al(2008{\natexlab{b}})Manning, Raghavan, and
  Sch{\"u}tze}]{manning2008text}
Manning CD, Raghavan P, Sch{\"u}tze H (2008{\natexlab{b}}) \emph{Text
  classification and Naive Bayes}, Cambridge University Press, p 234–265.
  \doi{10.1017/CBO9780511809071.014}

\bibitem[{McCreadie et~al(2013)McCreadie, Macdonald, Ounis, Osborne, and
  Petrovic}]{mccreadie2013scalable}
McCreadie R, Macdonald C, Ounis I, Osborne M, Petrovic S (2013) Scalable
  distributed event detection for twitter. In: 2013 IEEE international
  conference on big data, IEEE, pp 543--549

\bibitem[{McMinn et~al(2013)McMinn, Moshfeghi, and Jose}]{mcminn2013building}
McMinn AJ, Moshfeghi Y, Jose JM (2013) Building a large-scale corpus for
  evaluating event detection on twitter. In: Proceedings of the 22nd ACM
  international conference on Information \& Knowledge Management, pp 409--418

\bibitem[{Mikolov et~al(2010)Mikolov, Karafi{\'a}t, Burget, {\v{C}}ernock{\`y},
  and Khudanpur}]{mikolov2010recurrent}
Mikolov T, Karafi{\'a}t M, Burget L, {\v{C}}ernock{\`y} J, Khudanpur S (2010)
  Recurrent neural network based language model. In: Eleventh annual conference
  of the international speech communication association

\bibitem[{Mikolov et~al(2013{\natexlab{a}})Mikolov, Chen, Corrado, and
  Dean}]{mikolov2013efficient}
Mikolov T, Chen K, Corrado G, Dean J (2013{\natexlab{a}}) Efficient estimation
  of word representations in vector space. \emph{arXiv preprint arXiv:13013781}

\bibitem[{Mikolov et~al(2013{\natexlab{b}})Mikolov, Sutskever, Chen, Corrado,
  and Dean}]{mikolov2013distributed}
Mikolov T, Sutskever I, Chen K, Corrado GS, Dean J (2013{\natexlab{b}})
  Distributed representations of words and phrases and their compositionality.
  In: Advances in neural information processing systems, pp 3111--3119

\bibitem[{Morabia et~al(2019)Morabia, Murthy, Malapati, and
  Samant}]{morabia2019sedtwik}
Morabia K, Murthy NLB, Malapati A, Samant S (2019) Sedtwik: Segmentation-based
  event detection from tweets using wikipedia. In: Proceedings of the 2019
  Conference of the North American Chapter of the Association for Computational
  Linguistics: Student Research Workshop, pp 77--85

\bibitem[{M{\"u}llner(2011)}]{mullner2011modern}
M{\"u}llner D (2011) Modern hierarchical, agglomerative clustering algorithms.
  \emph{arXiv preprint arXiv:11092378}

\bibitem[{Nguyen et~al(2019)Nguyen, Ngo, Vo, and Cao}]{nguyen2019hot}
Nguyen S, Ngo B, Vo C, Cao T (2019) Hot topic detection on twitter data streams
  with incremental clustering using named entities and central centroids. In:
  2019 IEEE-RIVF International Conference on Computing and Communication
  Technologies (RIVF), IEEE, pp 1--6

\bibitem[{Nur'Aini et~al(2015)Nur'Aini, Najahaty, Hidayati, Murfi, and
  Nurrohmah}]{nur2015combination}
Nur'Aini K, Najahaty I, Hidayati L, Murfi H, Nurrohmah S (2015) Combination of
  singular value decomposition and k-means clustering methods for topic
  detection on twitter. In: 2015 International Conference on Advanced Computer
  Science and Information Systems (ICACSIS), IEEE, pp 123--128

\bibitem[{Parikh and Karlapalem(2013)}]{parikh2013events}
Parikh R, Karlapalem K (2013) Et: events from tweets. In: Proceedings of the
  22nd international conference on world wide web, pp 613--620

\bibitem[{Pennington et~al(2014)Pennington, Socher, and
  Manning}]{pennington2014glove}
Pennington J, Socher R, Manning C (2014) Glove: Global vectors for word
  representation. In: Proceedings of the 2014 conference on empirical methods
  in natural language processing (EMNLP), pp 1532--1543

\bibitem[{Pollard and Sag(1987)}]{sag1987information}
Pollard C, Sag IA (1987) \emph{Information-based syntax and semantics}.
  Cambridge university press

\bibitem[{Roux(2018)}]{roux2018comparative}
Roux M (2018) A comparative study of divisive and agglomerative hierarchical
  clustering algorithms. \emph{Journal of Classification} 35(2):345--366

\bibitem[{Sanh et~al(2019)Sanh, Debut, Chaumond, and Wolf}]{sanh2019distilbert}
Sanh V, Debut L, Chaumond J, Wolf T (2019) Distilbert, a distilled version of
  bert: smaller, faster, cheaper and lighter. \emph{arXiv preprint
  arXiv:191001108}

\bibitem[{Sayyadi et~al(2009)Sayyadi, Hurst, and Maykov}]{sayyadi2009event}
Sayyadi H, Hurst M, Maykov A (2009) Event detection and tracking in social
  streams. In: Third International AAAI Conference on Weblogs and Social Media

\bibitem[{Schakel and Wilson(2015)}]{schakel2015measuring}
Schakel AM, Wilson BJ (2015) Measuring word significance using distributed
  representations of words. \emph{arXiv preprint arXiv:150802297}

\bibitem[{Schinas et~al(2015)Schinas, Papadopoulos, Petkos, Kompatsiaris, and
  Mitkas}]{schinas2015multimodal}
Schinas M, Papadopoulos S, Petkos G, Kompatsiaris Y, Mitkas PA (2015)
  Multimodal graph-based event detection and summarization in social media
  streams. In: Proceedings of the 23rd ACM international conference on
  Multimedia, ACM, pp 189--192

\bibitem[{{\v{S}}krlj et~al(2020){\v{S}}krlj, Kralj, and
  Lavra{\v{c}}}]{vskrlj2020embedding}
{\v{S}}krlj B, Kralj J, Lavra{\v{c}} N (2020) Embedding-based silhouette
  community detection. \emph{Machine Learning} 109(11):2161--2193

\bibitem[{Small and Medsker(2014)}]{small2014review}
Small SG, Medsker L (2014) Review of information extraction technologies and
  applications. \emph{Neural computing and applications} 25(3-4):533--548

\bibitem[{Tsai(2009)}]{tsai2009mining}
Tsai PS (2009) Mining frequent itemsets in data streams using the weighted
  sliding window model. \emph{Expert Systems with Applications}
  36(9):11617--11625

\bibitem[{Van~Oorschot et~al(2012)Van~Oorschot, Van~Erp, and
  Dijkshoorn}]{van2012automatic}
Van~Oorschot G, Van~Erp M, Dijkshoorn C (2012) Automatic extraction of soccer
  game events from twitter. In: Proceedings of the Workhop on Detection,
  Representation, and Exploitation of Events in the Semantic Web (DeRiVE 2012),
  pp 21--30

\bibitem[{Weiler et~al(2017)Weiler, Grossniklaus, and
  Scholl}]{weiler2017survey}
Weiler A, Grossniklaus M, Scholl MH (2017) Survey and experimental analysis of
  event detection techniques for twitter. \emph{The Computer Journal}
  60(3):329--346

\bibitem[{Wolf et~al(2019)Wolf, Debut, Sanh, Chaumond, Delangue, Moi, Cistac,
  Rault, Louf, Funtowicz et~al}]{wolf2019huggingface}
Wolf T, Debut L, Sanh V, Chaumond J, Delangue C, Moi A, Cistac P, Rault T, Louf
  R, Funtowicz M, et~al (2019) Huggingface's transformers: State-of-the-art
  natural language processing. \emph{ArXiv} pp arXiv--1910

\bibitem[{Xie et~al(2016)Xie, Zhu, Jiang, Lim, and Wang}]{xie2016topicsketch}
Xie W, Zhu F, Jiang J, Lim EP, Wang K (2016) Topicsketch: Real-time bursty
  topic detection from twitter. \emph{IEEE Transactions on Knowledge and Data
  Engineering} 28(8):2216--2229

\bibitem[{Xu et~al(2007)Xu, Yuruk, Feng, and Schweiger}]{xu2007scan}
Xu X, Yuruk N, Feng Z, Schweiger TA (2007) Scan: a structural clustering
  algorithm for networks. In: Proceedings of the 13th ACM SIGKDD international
  conference on Knowledge discovery and data mining, ACM, pp 824--833

\bibitem[{Yang et~al(2019)Yang, Xie, Lin, Li, Tan, Xiong, Li, and
  Lin}]{yang-etal-2019-end}
Yang W, Xie Y, Lin A, Li X, Tan L, Xiong K, Li M, Lin J (2019) End-to-end
  open-domain question answering with {BERT}serini. In: Proceedings of the 2019
  Conference of the North {A}merican Chapter of the Association for
  Computational Linguistics (Demonstrations), Association for Computational
  Linguistics, Minneapolis, Minnesota, pp 72--77, \doi{10.18653/v1/N19-4013},
  \urlprefix\url{https://www.aclweb.org/anthology/N19-4013}

\bibitem[{Yilmaz and Toklu(2020)}]{yilmaz2020deep}
Yilmaz S, Toklu S (2020) A deep learning analysis on question classification
  task using word2vec representations. \emph{Neural Computing and Applications}
  32:2909--2928

\bibitem[{Zhang et~al(2019)Zhang, Liu, and Gulla}]{zhang2019dynamic}
Zhang L, Liu P, Gulla JA (2019) Dynamic attention-integrated neural network for
  session-based news recommendation. \emph{Machine Learning} 108(10):1851--1875

\end{thebibliography}

\appendix
\section{Intermediate processing time} \label{appendix-intermidiate-time}
Even though we reported the execution time for the full process in the paper, we did an intermediate analysis to understand the complexity of each individual step. The obtained results on both data sets are summarised in Table \ref{tab:sub-processing-time-munliv} and \ref{tab:sub-processing-time-brexitvote}. In these tables, total time reports the time taken by whole corpus and average time reports the time taken by a single time window. Comparing the two data sets, MUNLIV has 58 2-minute time windows and Brexitvote has 12 30-minute time windows. 

As discussed in Section \ref{sub-sec:computational-complexity}, embedding learner and event window identifier are the complex components in Embed2Detect architecture which take a comparatively large proportion of the total execution time with the increase of data size. However, according to the obtained results, even with sequential processing, data can be processed in less time than the time taken for their generation. With parallel processing, the execution time can be further reduced to be more appropriate for real-time processing.

\begin{table*}
\caption{Embed2Detect intermediate processing time - MUNLIV}
\label{tab:sub-processing-time-munliv} 
\begin{tabular}{l|cc|cc}
\hline\noalign{\smallskip}
\bf \multirow{2}{*}{Step}  & \multicolumn{2}{c|}{\bf \makecell[c]{Execution Time(s)\\(1 worker)}} & \multicolumn{2}{c}{\bf \makecell[c]{Execution Time(s)\\(8 workers)}}\\
& Total & Average & Total & Average\\
\noalign{\smallskip}\hline\noalign{\smallskip}
Stream chunking & 66 & 1.158 & 64 & 1.123 \\
Embedding learning & 114 & 2 & 90 & 1.579 \\
Event window identification & 35 & 0.614 & 34 & 0.596 \\
Event word extraction & 0 & 0 & 0 & 0 \\
\noalign{\smallskip}\hline\noalign{\smallskip}
\bf Full process & 230 & 4.035 & 202 & 3.544 \\
\noalign{\smallskip}\hline\noalign{\smallskip}
\end{tabular}
\end{table*}

\begin{table*}
\caption{Embed2Detect intermediate processing time - BrexitVote}
\label{tab:sub-processing-time-brexitvote} 
\begin{tabular}{l|cc|cc}
\hline\noalign{\smallskip}
\bf \multirow{2}{*}{Step}  & \multicolumn{2}{c|}{\bf \makecell[c]{Execution Time(s)\\(1 worker)}} & \multicolumn{2}{c}{\bf \makecell[c]{Execution Time(s)\\(8 workers)}}\\
& Total & Average & Total & Average\\
\noalign{\smallskip}\hline\noalign{\smallskip}
Stream chunking & 28 & 2.545 & 27 & 2.455 \\
Embedding learning & 168 & 15.273 & 78 & 7.091 \\
Event window identification & 562 & 51.091 & 139 & 12.636 \\
Event word extraction & 29 & 2.636 & 29 & 2.636 \\
\noalign{\smallskip}\hline\noalign{\smallskip}
\bf Full process & 824 & 74.909 & 310 & 28.182 \\
\noalign{\smallskip}\hline\noalign{\smallskip}
\end{tabular}
\end{table*}

\end{document}